\documentclass[smallextended]{svjour3}       
\smartqed  
\usepackage[dvips]{graphicx}
\usepackage{mathptmx}      
\usepackage[usenames,dvipsnames]{color}
\usepackage{psfrag}

\newcommand{\farcs}{\hbox{\ensuremath{.\!\!^{\prime\prime}}}}
\newcommand{\farcm}{\hbox{\ensuremath{.\!\!^{\prime}}}}
\newcommand{\tfrac}[2]{\hbox{\ensuremath{\ ^{#1}\!\!/_{\!\!#2}}}}
\journalname{Experimental Astronomy}
\begin{document}

\title{The Herschel PACS photometer calibration}
\subtitle{A time dependent flux calibration for the PACS chopped point-source 
photometry AOT mode}

\titlerunning{The Herschel PACS photometer calibration: chopped point-source 
flux calibration}

\author{Markus Nielbock      \and
        Thomas M\"uller      \and
	Ulrich Klaas         \and
	Bruno Altieri        \and
        Zolt\'an Balog       \and
        Nicolas Billot       \and
        Hendrik Linz         \and
	Koryo Okumura        \and
	Miguel S\'anchez-Portal \and
	Marc Sauvage
}


\institute{M. Nielbock, Z. Balog, U. Klaas, H. Linz \at
           Max-Planck-Institut f\"ur Astronomie, K\"onigstuhl 17, D-69117
	   Heidelberg, Germany \\
           Tel.: +49 6221 528-445\\
           \email{nielbock@mpia.de}
           \and
	   B. Altieri, M. S\'anchez-Portal \at
	   European Space Astronomy Centre (ESAC)/ESA, Villanueva de la
	   Ca\~nada, E-28691 Madrid, Spain
           \and
           N. Billot \at
           Instituto de Radio Astronom\'ia Milim\'etrica, Avenida Divina
           Pastora, 7, Local 20, 18012 Granada, Spain
           \and
           Th. M\"uller \at
	   Max-Planck-Institut f\"ur extraterrestrische Physik,
	   Gie{\ss}enbachstra{\ss}e, D-85741 Garching, Germany
           \and
	   K. Okumura, M. Sauvage \at
	   Laboratoire AIM, CEA, Universit\'e Paris Diderot, IRFU/Service
	   d'Astrophysique, Bat. 709, 91191 Gif-sur-Yvette, France
}

\date{Received: date / Accepted: date}

\maketitle

\begin{abstract}
We present a flux calibration scheme for the PACS chopped point-source 
photometry observing mode based on the photometry of five stellar standard 
sources. This mode was used for science observations only early in the mission. 
Later, it was only used for pointing and flux calibration measurements. Its 
calibration turns this type of observation into fully validated data products 
in the Herschel Science Archive. Systematic differences in calibration with 
regard to the principal photometer observation mode, the scan map, are derived 
and amount to $5-6$\%. An empirical method to calibrate out an apparent 
response drift during the first 300 Operational Days is presented. The relative 
photometric calibration accuracy (repeatability) is as good as 1\% in the blue 
and green band and up to 5\% in the red band. Like for the scan map mode, 
inconsistencies among the stellar calibration models become visible and amount 
to 2\% for the five standard stars used. The absolute calibration accuracy is 
therefore mainly limited by the model uncertainty, which is 5\% for all three 
bands.

\end{abstract}

\begin{table}
\caption{Specifications of the five prime flux standards used for calibrating 
the PACS photometer response.}
\label{tab:prime}
\begin{tabular}{lrllcrrr}
\hline\noalign{\smallskip}
&&\multicolumn{2}{c}{ICRS coordinates (J2000)}& Spectral & 
\multicolumn{3}{c}{Model flux [mJy]} \\ 
\multicolumn{1}{c}{ID} & \multicolumn{1}{c}{HIP} & \multicolumn{1}{c}{RA} &
\multicolumn{1}{c}{Dec} & \multicolumn{1}{c}{Type} & \multicolumn{1}{c}{70
$\mu$m} & \multicolumn{1}{c}{100 $\mu$m} & \multicolumn{1}{c}{160 $\mu$m} \\
\noalign{\smallskip}\hline\noalign{\smallskip}
$\alpha$~Boo & 69673 & 14:15:39.67207 & +19:10:56.6730 & K1.5III  & 15434 & 
7509 & 2891 \\
$\alpha$~Cet & 14135 & 03:02:16.77307 & +04:05:23.0596 & M1.5IIIa &  4889 & 
2393 &  928 \\
$\alpha$~Tau & 21421 & 04:35:55.23907 & +16:30:33.4885 & K5III    & 14131 & 
6909 & 2677 \\
$\beta$~And  &  5447 & 01:09:43.92388 & +35:37:14.0075 & M0III    &  5594 & 
2737 & 1062 \\
$\gamma$~Dra & 87833 & 17:56:36.36988 & +51:29:20.0242 & K5III    &  3283 & 
1604 &  621 \\
\noalign{\smallskip}\hline
\noalign{\smallskip}
\end{tabular}
\textbf{Note.} The coordinates are taken from \cite{hipparcos07}. The model
fluxes are based on \cite{dehaes11}. 
\end{table}

\section{Introduction}
\label{intro}
The Photodetector Array Camera and Spectrograph (PACS\footnote{PACS has been 
developed by a consortium of institutes led by MPE (Germany) and including UVIE 
(Austria); KU Leuven, CSL, IMEC (Belgium); CEA, LAM (France); MPIA (Germany); 
INAF-IFSI/OAA/OAP/OAT, LENS, SISSA (Italy); IAC (Spain). This development has 
been supported by the funding agencies BMVIT (Austria), ESA-PRODEX (Belgium), 
CEA/CNES (France), DLR (Germany), ASI/INAF (Italy), and CICYT/MCYT (Spain).}) 
\cite{pacs} on board the Herschel Space Observatory~\cite{herschel} provides 
photometric imaging and integral field spectroscopy capabilities for the 
far-infrared (FIR) wavelength regime. The PACS photometer unit is a dual band 
imaging camera. It permits the simultaneous observation in two bands, either in 
the combination 70~$\mu$m/160~$\mu$m or 100~$\mu$m/160~$\mu$m, where 70 or 
100~$\mu$m bands are selected by a filter wheel. The camera contains two 
bolometer detector arrays with $64\times32$ pixels (blue array) and 
$32\times16$ pixels (red array) providing an instantaneous field-of-view of 
$3\farcm5\times1\farcm75$. The detector arrays are made of 8 and 2 filled 
bolometer matrices with $16\times16$ pixels each, respectively. They operate at 
$\sim 285$~mK provided by a $^3$He sorption cooler with a hold time of 57.8~h, 
if the bolometers are biased all the time after the cooler recycling.

\begin{figure}
\centering
\includegraphics[height=2.3cm]{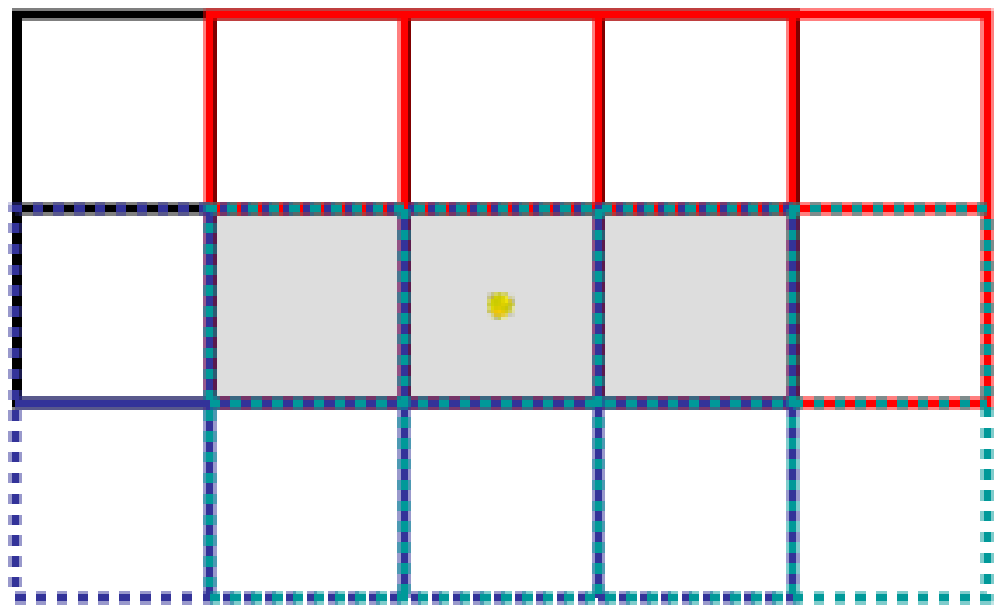}
\hspace*{1cm}
\includegraphics[height=2.3cm]{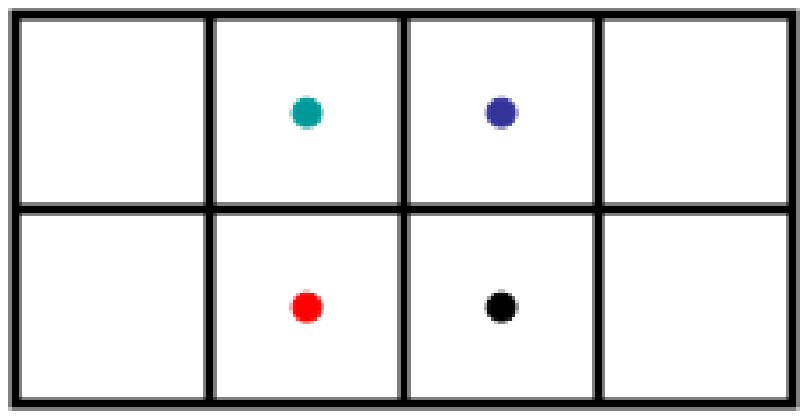}
\caption{
Illustration of the detector footprint of the blue detector array on the sky
(left) and the chop-nod source pattern (right) produced on the detector during
the execution of the point-source photometry AOT that offsets the observed
direction by about $50''$ in perpendicular directions, both for chopping and
nodding. The colours reflect the four combinations of the nodding and chopping
positions attained during the observing sequence:
\textcolor{black}{$\bullet$} nod 1 chop A,
\textcolor{red}{$\bullet$} nod 1 chop B, 
\textcolor{Blue}{$\bullet$} nod 2 chop A,
\textcolor{Aquamarine}{$\bullet$} nod 2 chop B.
The detector array with a field-of-view of $3\farcm5\times1\farcm75$
consists of eight individual sub-matrices.}
\label{fig:obsmode}
\end{figure}

While the principal science observation mode with the PACS photometer is the 
scan map with the telescope scanning along parallel legs covering the map 
area\footnote{http://herschel.esac.esa.int/Docs/PACS/html/pacs\_om.html}, 
another photometer observing mode was maintained throughout the mission: the 
chopped point-source photometry Astronomical Observing Template 
(AOT)\footnote{http://herschel.esac.esa.int/twiki/pub/Public/PacsCalibrationWeb/PhotMiniScan\_ReleaseNote\_20101112.pdf}.
{
For a general description of the differential measurement technique by chopping,
see \cite{emerson94}. The chopping technique has been used in space-borne FIR
instrumentation like ISOPHOT \cite{isophot} and ground-based mid-infrared
cameras like TIMMI2 \cite{timmi2} and VISIR \cite{visir}. Inside PACS, it is
achieved by using
}
the instrument internal focal plane 
chopper to provide signal modulation by on-array chopping (see 
Fig.~\ref{fig:obsmode}). This restricts the useful field-of-view to roughly 
$50''$, hence is only useful for point or very compact source photometry. It was 
used for science observations early in the mission and later only as calibration 
mode for 

\begin{enumerate}
\item Observatory pointing calibration, which allowed a more accurate 
          source position determination than scan mapping,
\item a low cost independent flux calibration check on the same celestial 
          standards as used for the scan map calibration.
\end{enumerate}

{
In total, 2200 observations that sum up to 200 hours of Herschel observing
time have been executed in this mode, adding a wealth of astronomical data to
the Herschel Science Archive (HSA). 16 hours are from scientific observing
programmes, while 184 hours are spent on calibration observations of point
sources like stars, asteroids and planets.
}
It is therefore the intention of this paper to make this
type of observation fully validated data products by providing a consistent
flux calibration scheme. For an overview of the basic calibration strategy we 
refer to the analogous steps of the scan map flux calibration~\cite{balog13}.
Here we concentrate on working out and characterising the specific properties
of this mode.
 
\section{Observations}
\label{sec:obs}
Altogether, 137 observations (see Tab.~\ref{tab:obslist}) of the five PACS prime
standard stars (see Tab.~\ref{tab:prime}) were obtained using the chopped
point-source photometry AOT.
{
The stellar models\footnote{Model spectra are available at:
ftp://ftp.ster.kuleuven.be/dist/pacs/calsources} are based on  \cite{dehaes11}.
For a more detailed discussion see \cite{balog13}.
}
In the case of PACS, chopping is done with a focal plane chopping mirror
\cite{krause00} operating at a frequency of 1.25~Hz that produces a displacement
of $\sim 50^{\prime\prime}$ on the sky along the detector y-axis. Synchronised
with the chopper movement, the data are read out with a rate of 40~Hz, but
averaged to a 10~Hz resolution on board (see Fig.~\ref{fig:chopsignal}). Nodding
is achieved by offsetting the telescope by the same amount along the detector
z-axis. On each nod-position, the chopper executes $3\times25$ cycles. The three
sets of chopping cycles are either on the same array position (no dithering) or
on 3 different array positions (dither option). If dithering is applied, the
chopper pattern is displaced by $2\tfrac{2}{3}$ blue or by $1\tfrac{1}{3}$ red
detector pixels along the detector y-axis. The detector footprint on sky without
dithering as well as the resulting target pattern on the detector is shown in
Fig.~\ref{fig:obsmode}. For details see the PACS Observer's
Manual\footnote{http://herschel.esac.esa.int/Docs/PACS/html/pacs\_om.html}. The
observation of the science target is always preceded by a chopped measurement on
the internal calibration sources, known as the calibration block, or brief,
calblock. This information is not used for calibrating the detector response
drifts during processing, but serves as a basis for long term trend corrections.
A detailed analysis is given in \cite{moor13}.

The observations were always set up in a very similar manner, e.~g.~fixed to a
single repetition which leads to identical on-target times of 124~s and a
duration of 152~s for the Astronomical Observing Request (AOR) including
overheads but omitting the initial slew. Apart from two exceptional cases,
dithering was used. However, we did not see any significant difference in terms
of measured target flux.
{
The gain parameter was set to ``high'' for nearly all observations. However,
changing the gain parameter has no influence on the photometry of the selected
targets.
}
The number
of observations per object and spectral band is given in Tab.~\ref{tab:obsstat}.

\begin{table}
\caption{Numbers of observations per object and spectral band. $\gamma$~Dra is
the designated response stability monitoring target. Thus, it was observed more
frequently than the other four objects.} 
\label{tab:obsstat}
\begin{tabular}{lrrr}
\hline\noalign{\smallskip}
\multicolumn{1}{c}{Object} & \multicolumn{1}{c}{70 $\mu$m} & 
\multicolumn{1}{c}{100 $\mu$m} & \multicolumn{1}{c}{160 $\mu$m} \\
\noalign{\smallskip}\hline\noalign{\smallskip}
$\alpha$~Boo &  8 &  7 & 15 \\ 
$\alpha$~Cet & 10 &  7 & 17 \\
$\alpha$~Tau & 13 & 10 & 23 \\
$\beta$~And  &  6 &  6 & 12 \\
$\gamma$~Dra & 61 &  9 & 70 \\
\noalign{\smallskip}\hline
\end{tabular}
\end{table}

Please note that on operational day (OD) 1375, one of the two red detector 
matrices
failed and never recovered. As of PACS calibration
version\footnote{http://herschel.esac.esa.int/twiki/bin/view/Public/PacsCalTreeHistory}
48, this unusable matrix is
flagged automatically during the data processing. As a result, all observations
at 160~$\mu$m obtained since then are affected by a reduced spatial coverage and
consequently by a degraded sensitivity.

\section{Data processing}
\label{sec:processing}

The data have been processed in a straightforward manner that is effectively
very similar to the standard
pipeline\footnote{http://herschel.esac.esa.int/hcss-doc-10.0/index.jsp\#pacs\_phot:PdrgP.Chp.3.chopnod} \cite{wieprecht09} (version 9.1 and later) provided by the Herschel Science 
Centre (HSC) of the
European Space Agency (ESA) via the HSA. In particular, it comprises 

\begin{itemize}
	\item flagging of outliers (bad pixels, electronic saturation and
		crosstalk, data recorded during chopper transitions)
	\item averaging data per chopper plateau (first readout is discarded due
		to slow detector response)
	\item producing differential signals per chop cycle
	\item response calibration and flat fielding
	\item sorting for and averaging per dither position
	\item deglitching via sigma clipping algorithm
	\item averaging data per nod position and subtracting
	\item producing map via shift-and-add algorithm
\end{itemize}

\begin{figure}
\centering
\includegraphics[width=0.60\textwidth]{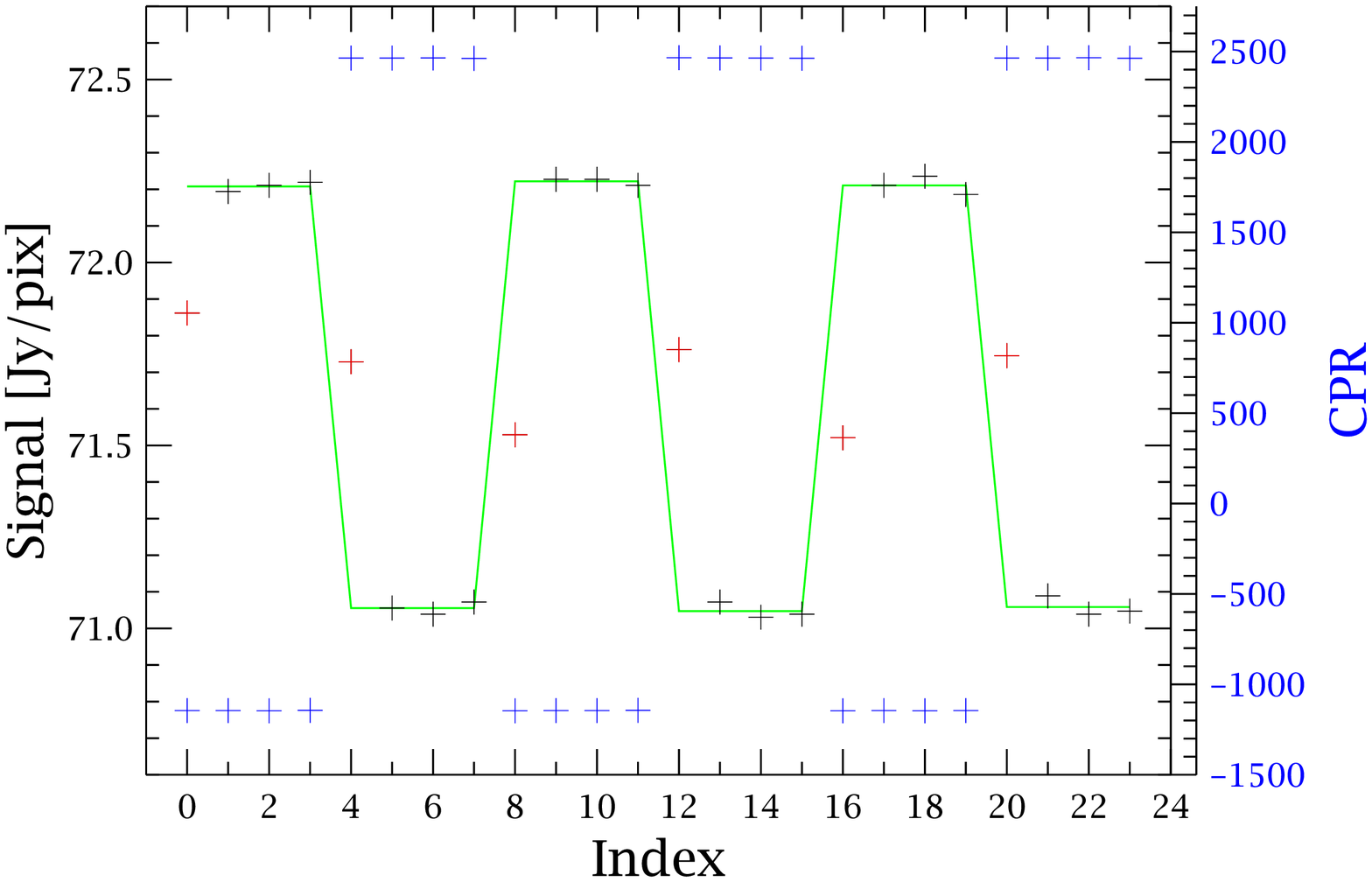}
\caption{Detailed view of the detector signals for a single bolometer pixel
during a sequence of three chopping cycles. The detector signals are in black,
the averaged value per chopper plateau is shown as a green line. The red crosses
denote detector signals that have been discarded. Although the chopper reaches
the commanded position (blue crosses, in digital readout units), the first
detector signal of a chopper plateau is too low due to the inertia of the 
response.}
\label{fig:chopsignal}
\end{figure}

{
The first two steps are visualised in Fig.~\ref{fig:chopsignal}. The blue
crosses show how the chopper position alternates between two values given in
digital readout units. The red and
black crosses represent the detector signal measured at those chopper positions.
The first readout obtained at a new chopper position does not provide the true
full signal (red crosses) due to a detector response lag phenomenon. Therefore,
those data are flagged and removed from the subsequent processing. The green
line depicts the detector signal after averaging over a given chopper plateau.
}

The unreleased PACS calibration version 53 was used, which for most PACS
photometer aspects is identical to versions 41 (SPG 9.1) and 48 (SPG 10), which
is the latest publicly released version. Therefore, the calibration set is
almost identical to what was used for the Standard Pipeline Generation (SPG). In
particular, the standard response calibration (FM,7) was applied which is based 
on scan map observations of the five prime calibration stars.

In addition, we have included an updated detector pixel table containing the
corner positions for each pixel, which reflects the optical distortions on the
sky. Another improvement is a correction for response changes that are caused
by temperature variations of the bolometer $^3$He cooler \cite{balog13,moor13}.
For the dataset presented here, the correction changes the flux levels by
$(+0.6 \pm 0.6)$\% at 70~$\mu$m, 
$(+0.7 \pm 0.8)$\% at 100~$\mu$m, and  
$(+0.5 \pm 0.5)$\% at 160~$\mu$m.
Those corrections will be made available in future releases of the Herschel
Interactive Processing Environment (HIPE) \cite{hipe}. A non-linearity 
correction was not
necessary, because all the five standard stars are too faint to cause any
detector non-linearity effects. The thresholds of non-linearity effects caused
by point source fluxes are approximately 50, 70, and 40~Jy at 70, 100, and
160~$\mu$m, respectively.

We have also used corrected pointing products as provided by the
HSC\footnote{http://herschel.esac.esa.int/twiki/bin/view/Public/HowToUseImprovedPointingProducts}.
They provide focal length field distortion corrections of the star tracker
camera for ODs 320 until 761 and hence improve the telescope pointing
information during data processing. Revised pointing products for earlier ODs
are not yet available. For observations obtained at later ODs, the correct focal
length was used for the pointing calculations on-board.

We used a pixel scaling for creating the final point-source maps that differs
from the SPG settings. It was chosen to appropriately sample the target point
spread function (PSF) by covering the nominal full width at half maximum (FWHM)
with five map pixels. The resulting map pixel scales are $1\farcs1$, $1\farcs4$,
and $2\farcs1$ at 70, 100, and 160~$\mu$m, respectively.

\begin{figure}
\centering
\includegraphics[width=0.60\textwidth]{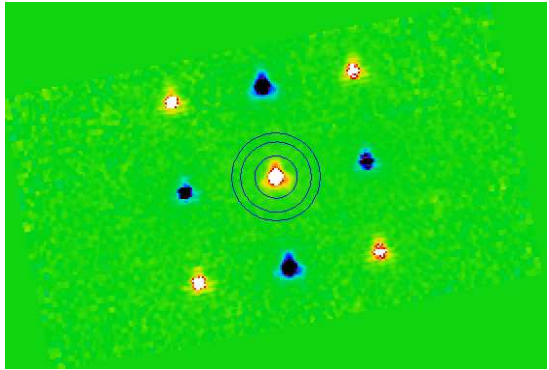}
\caption{Point-source map of $\gamma$~Dra obtained at 70~$\mu$m.
It is produced by a shift-and-add technique that co-adds the four images of the
source that are produced by the chop-nod observing method (see
Fig.~\ref{fig:obsmode}). As a result, only the central image of the nine
must be used for photometry and further analysis. The triangular shape of the
source reflects the typical PSF of a point source. 
The aperture ($12''$) and sky annulus used for the photometry are superimposed.
The FOV of the co-added map is $4\farcm7\times2\farcm7$, but the useful area of
the central target image is given by the chop and the nod throw, i.e.
approximately $50''$.}
\label{fig:map}
\end{figure}

\section{Photometry}
\label{sec:photometry}

The \texttt{annularSkyAperturePhotometry} task of HIPE was used to determine the
source flux from the point-source map that is produced by applying a
shift-and-add algorithm. For observations obtained at 70 and 100~$\mu$m, we set
the aperture radius to $12^{\prime\prime}$, the inner sky annulus to
$20^{\prime\prime}$ and the outer sky annulus to $25^{\prime\prime}$. For
observations in the 160~$\mu$m band, we selected an aperture radius of
$22^{\prime\prime}$, while the sky annulus ranged from $24^{\prime\prime}$ to
$28^{\prime\prime}$. The maximum radius is constrained by the size of the inner
patch of the final map (Fig.~\ref{fig:map}) that contains the actual photometric
information. The sky range was only used to estimate the photometric
uncertainty. A background offset was not subtracted, because the chop-nod
observing technique already provides background subtracted maps for single and
isolated point-sources. We set the option \texttt{centroid} to \texttt{false}
and extracted the central source positions from 2D Gaussian fits beforehand
that were fed into the photometry algorithm.

An aperture correction was applied to account for the point-source flux outside
the aperture via the \texttt{photApertureCorrectionPointSource} task. It
is strictly only valid for scan map observations that produce PSFs
that are slightly different from the chop-nod observations. However, the
apertures are chosen large enough so that the differences are insignificant and
the source flux is reliably recovered. The correction factors by which the
measured point source fluxes are divided are 0.802, 0.776, and 0.817 at 70, 100,
and 160~$\mu$m, respectively.

{
The colour correction values for the fiducial stars (1.016, 1.033, 1.074) were
derived from a 4000 K black body. We repeated now the calculation using the
official model template files (see Sect.~\ref{sec:obs}) and could confirm the tabulated
values in the blue and red band. In the green band we found a difference of
0.1\% (1.034) when using the full stellar templates instead of a 4000~K black
body. No difference between the K and M-giants are seen on the per mille level.
In the final error budget, this small deviation can be neglected and the colour
correction is not contributing to the systematic differences between K and
M-giants seen in the calibrated flux densities of the 5 fiducial stars.
}

\begin{table}
\caption{Results of the photometry of the five prime standard stars. The quoted
numbers are the mean values of the flux ratios from OD~300 onwards. When
combining more than one object, the mean is weighted by the number of objects.}
\label{tab:photstat}
\begin{tabular}{lccc}
\hline\noalign{\smallskip}
& \multicolumn{3}{c}{$\left< F_\mathrm{obs}/F_\mathrm{model} 
\right>_{\geq\mathrm{OD~300}}$} \\
\multicolumn{1}{c}{Object(s)} & \multicolumn{1}{c}{70 $\mu$m} & 
\multicolumn{1}{c}{100 $\mu$m} & \multicolumn{1}{c}{160 $\mu$m} \\
\noalign{\smallskip}\hline\noalign{\smallskip}
$\alpha$~Boo & $0.928 \pm 0.011$ & $0.942 \pm 0.005$ & $0.958 \pm 0.009$ \\ 
$\alpha$~Cet & $0.949 \pm 0.006$ & $0.956 \pm 0.006$ & $0.991 \pm 0.031$ \\
$\alpha$~Tau & $0.912 \pm 0.008$ & $0.926 \pm 0.006$ & $0.933 \pm 0.011$ \\
$\beta$~And  & $0.959 \pm 0.011$ & $0.966 \pm 0.004$ & $0.938 \pm 0.012$ \\
$\gamma$~Dra & $0.927 \pm 0.009$ & $0.931 \pm 0.007$ & $0.948 \pm 0.046$ \\
\noalign{\smallskip}\hline
\noalign{\smallskip}
K giants     & $0.922 \pm 0.009$ & $0.933 \pm 0.008$ & $0.946 \pm 0.013$ \\
M giants     & $0.954 \pm 0.007$ & $0.961 \pm 0.007$ & $0.965 \pm 0.037$ \\
\noalign{\smallskip}\hline
\noalign{\smallskip}
all          & $0.935 \pm 0.019$ & $0.944 \pm 0.017$ & $0.954 \pm 0.023$ \\
\noalign{\smallskip}\hline
\end{tabular}
\end{table}

\begin{figure}
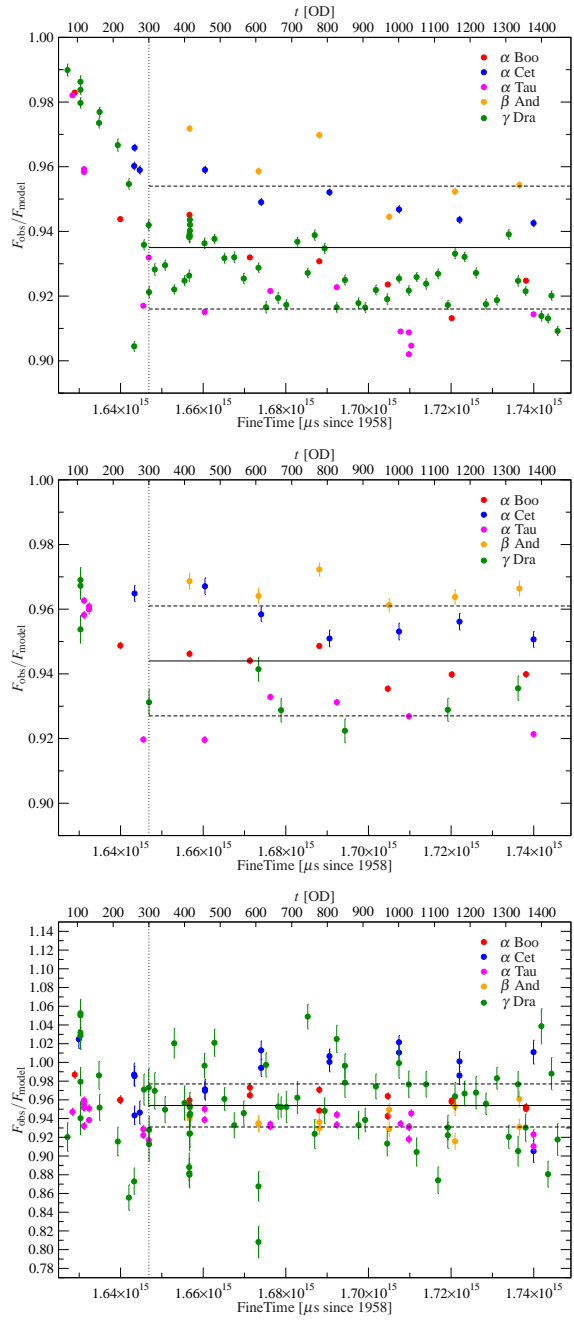

\centering
\psfrag{OD}[cc][][0.7][0]{$t$ [OD]}
\psfrag{FT}[cc][][0.7][0]{FineTime [$\mu$s since 1958]}
\psfrag{F}[cc][][0.7][0]{$F_\mathrm{obs}/F_\mathrm{model}$}
\psfrag{aBoo}[cc][][0.7][0]{$\alpha$~Boo}
\psfrag{aCet}[cc][][0.7][0]{$\alpha$~Cet}
\psfrag{aTau}[cc][][0.7][0]{$\alpha$~Tau}
\psfrag{bAnd}[cc][][0.7][0]{$\beta$~And}
\psfrag{gDra}[cc][][0.7][0]{$\gamma$~Dra}
\includegraphics[width=0.62\textwidth]{5FiducialChopNodModelRatiosTime_70.eps}

\medskip
\includegraphics[width=0.62\textwidth]{5FiducialChopNodModelRatiosTime_100.eps}

\medskip
\includegraphics[width=0.62\textwidth]{5FiducialChopNodModelRatiosTime_160.eps}
\caption{Photometric results of the stellar calibration sources at 70, 100, and
160~$\mu$m (top to bottom) vs. time. The time scales are given in
\texttt{FineTime} units, i.~e.~microseconds since 1 January 1958, and in
operational days (OD) since launch. The flux scale is given in ratios between
the model and the observed source fluxes.  The horizontal lines depict the mean
and the rms of the averaged photometric results for OD $\geq 300$ obtained for
the three wave bands.}
\label{fig:photometry}
\end{figure}

The photometric results obtained from the three PACS bands are given in
Fig.~\ref{fig:photometry}. For comparison, we show the ratios between the
modelled and observed source fluxes (see Tab.~\ref{tab:prime}). The error bars
represent the photometric uncertainties as they are calculated by the aperture
photometry algorithm. Note that the true photometric uncertainties can be
considerably higher, because the maps suffer from correlated noise. Empirical
investigations imply that for the chosen pixel sizes the true photometric
uncertainty can be higher by up to a factor of 6, 4, and 8 at 70, 100, and
160~$\mu$m, respectively. For the time scale, we provide both the elapsed time
since launch as ODs and the Herschel internal canonical \texttt{FineTime} format
as microseconds elapsed since 1 January 1958\footnote{This is the defined zero
point of the International Atomic Time (TAI) standard.}. For the subsequent
analysis, we only use \texttt{FineTime}, because the ODs have variable lengths.

\section{Results}
\label{sec:results}
Several results are obvious from the measurements presented in 
Fig.~\ref{fig:photometry}.

\begin{enumerate}
	\item There is an initial decline in response visible until OD~300, most
		 obviously witnessed in the 70~$\mu$m band.
	\item After OD~300, the flux ratios remain fairly constant.
	\item There is an apparent offset between the K giant stars
		($\alpha$~Boo, $\alpha$~Tau, $\gamma$~Dra) and the M giant stars
		($\alpha$~Cet, $\beta$~And) at 70~$\mu$m and 100~$\mu$m
                {
                of about 3\%},
                which is
		pointing to systematic uncertainties in the underlying stellar
		models that are quoted to be accurate to 5\% \cite{dehaes11}.
                {
                See also \cite{balog13} for a comparison with the PACS scan map
                calibration.}
\end{enumerate}

\begin{figure}
\centering
\psfrag{F}[cc][][0.9][0]{$F_\mathrm{model}$ [Jy]}
\psfrag{rms}[cc][cc][0.9][0]{rms [Jy]}
\psfrag{70}[Bl][bc][0.9][0]{70~$\mu$m}
\psfrag{100}[Bl][bl][0.9][0]{100~$\mu$m}
\psfrag{160}[Bl][bl][0.9][0]{160~$\mu$m}
\includegraphics[width=0.65\textwidth]{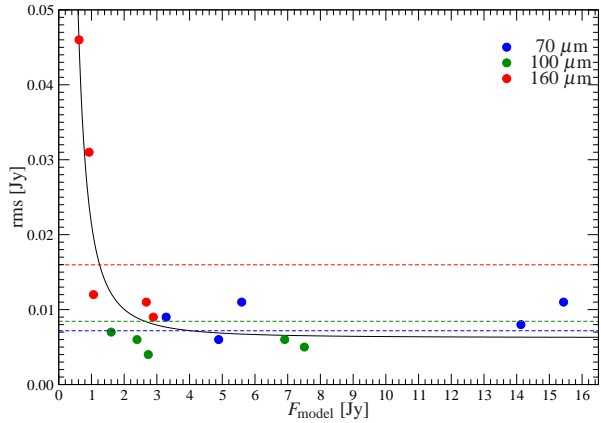}
\caption{Relation between the rms of the photometry for the individual objects 
and their
model flux in the three filter bands. The solid black line is a fit to the data
according to $\mathrm{rms} = c_1 \cdot F_\mathrm{model}^{-2}+c_2$. The dashed
horizontal lines represent the expected photometric uncertainties at the
corresponding three wave bands as predicted by the HSPOT observing planning 
tool.}
\label{fig:rms}
\end{figure}

\subsection{Observations obtained between OD~300 and the end of operations}
\label{sec:results_late}

Measurements after OD~300 yield results according to the statistics shown in
Tab.~\ref{tab:photstat}. Using those as a reference for comparing with the flux
calibration based on scan maps, the measured source fluxes obtained from
chop-nod observations are on average lower by 6.5\%, 5.6\% and 4.6\% at 70, 100,
and 160~$\mu$m, respectively, with an uncertainty that is of the order of 2\%.
These numbers already provide reasonable flux correction factors for chop-nod
observations obtained from OD~300 onwards.

Note that the quoted uncertainties of the individual objects only reflect the
reproducibility of the photometry. The absolute calibration uncertainty is
dominated by the model uncertainties that amount to 5\%. For some of the
objects, the uncertainties are higher than for the others, particularly at
160~$\mu$m. To rule out intrinsic flux variations as the cause, we have analysed
the correlation between the model flux and the rms of the individual
object photometries. As shown in Fig.~\ref{fig:rms}, the increased scatter of 
the photometry of $\alpha$~Cet and $\gamma$~Dra at 160~$\mu$m anti-correlates 
with their model
flux. In addition, the results of the scan-map observations have a higher 
fidelity and
do not scatter that much \cite{balog13}. 
Therefore, the obtained uncertainty is not due to real flux variations, but can 
be
entirely explained by a reduced S/N of an individual measurement. This also
demonstrates that for relatively faint objects a photometric error prediction is
unreliable.

\begin{figure}
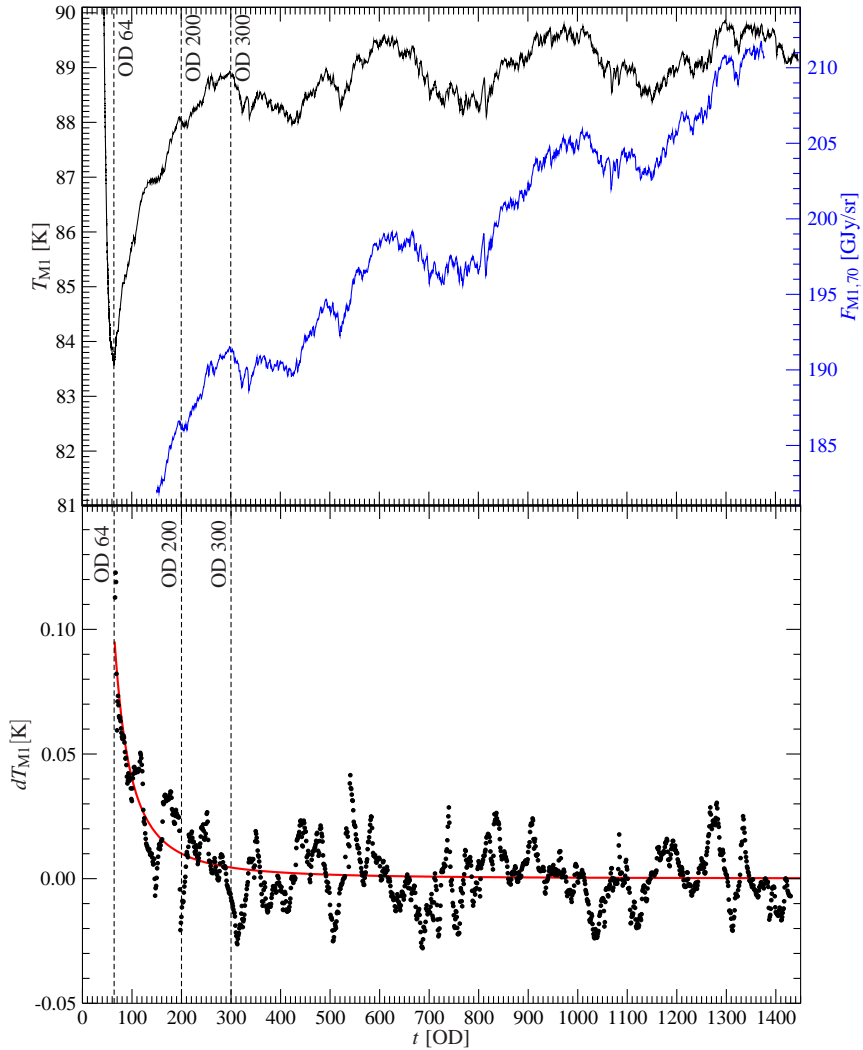

\centering
\psfrag{tod}[cc][][1.0][0]{$t$ [OD]}
\psfrag{tm1}[cc][][1.0][0]{$T_\mathrm{M1}$ [K]}
\psfrag{fm1}[cc][][1.0][0]{\color{blue}$F_\mathrm{M1,70}$ [GJy/sr]}
\psfrag{dT}[cc][][1.0][0]{$dT_\mathrm{M1} [\mathrm{K}]$}
\psfrag{od064}[][][1.0][0]{OD 64}
\psfrag{od200}[][][1.0][0]{OD 200}
\psfrag{od300}[][][1.0][0]{OD 300}
\includegraphics[scale=0.45]{M1TempFlux.eps}

\vspace*{-2mm}
\hspace*{-10.6mm}
\includegraphics[scale=0.45]{M1Temp_diff_psfrag.eps}
\caption{Evolution of the telescope main mirror temperature (black line, upper
panel), of the modelled flux density at 70~$\mu$m (blue line, upper panel) and
of the first derivative of the telescope main mirror temperature
(lower panel) during the Herschel mission. The solid red line is a fit to the
data according to $dT_\mathrm{M1} = c \cdot t^{-2}$.}
\label{fig:TM1}
\end{figure}

\subsection{Observations obtained until OD~300}
\label{sec:results_early}

\begin{figure}
\centering
\psfrag{F}[cc][][0.9][0]{$F_\mathrm{obs}/F_\mathrm{model}$}
\psfrag{GJy}[cc][][0.9][0]{$F_\mathrm{M1,70}$ [GJy/sr]}
\psfrag{gDra}[cc][][0.9][0]{$\gamma$~Dra}
\includegraphics[width=0.9\textwidth]{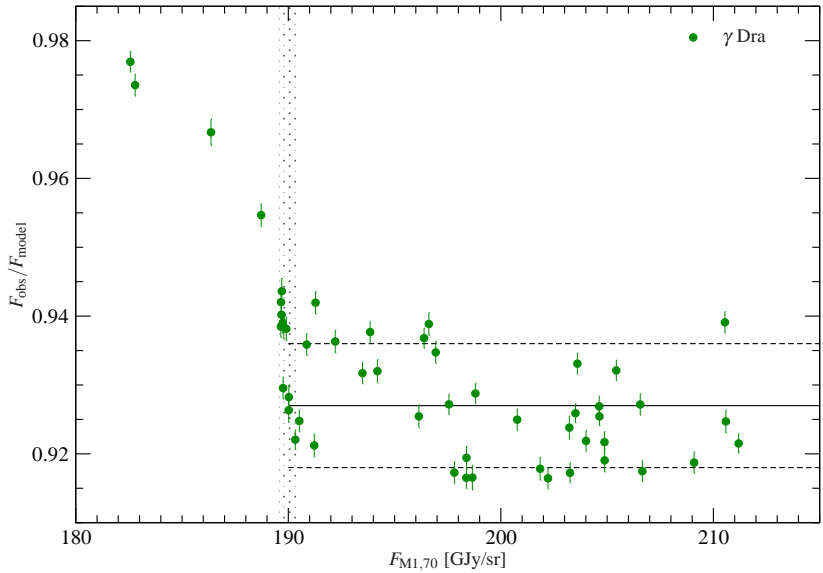}
\caption{Photometric results of $\gamma$~Dra vs. the corresponding modelled
telescope main mirror flux at 70~$\mu$m. The shaded area around 190~GJy/sr
represents the range of telescope fluxes attained around OD~300. The horizontal
lines represent the mean and the rms for ODs $\geq$ 300. All the data points
at lower flux values have been obtained on earlier ODs.}
\label{fig:FvsFM1}
\end{figure}

\begin{figure}
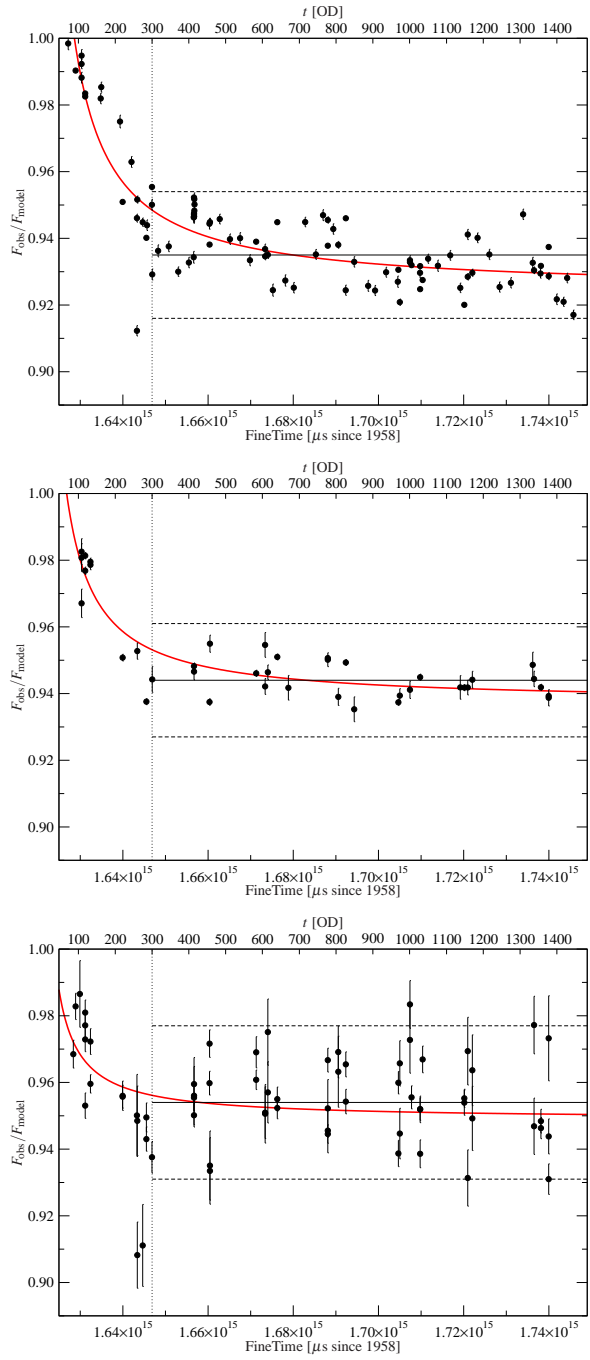

\centering
\psfrag{OD}[cc][][0.7][0]{$t$ [OD]}
\psfrag{FT}[cc][][0.7][0]{FineTime [$\mu$s since 1958]}
\psfrag{F}[cc][][0.7][0]{$F_\mathrm{obs}/F_\mathrm{model}$}
\psfrag{aBoo}[cc][][0.7][0]{$\alpha$~Boo}
\psfrag{aCet}[cc][][0.7][0]{$\alpha$~Cet}
\psfrag{aTau}[cc][][0.7][0]{$\alpha$~Tau}
\psfrag{bAnd}[cc][][0.7][0]{$\beta$~And}
\psfrag{gDra}[cc][][0.7][0]{$\gamma$~Dra}
\includegraphics[width=0.64\textwidth]{5FiducialChopNodNormalisedModelRatiosTime_70.eps}

\medskip
\includegraphics[width=0.64\textwidth]{5FiducialChopNodNormalisedModelRatiosTime_100.eps}

\medskip
\includegraphics[width=0.64\textwidth]{5FiducialChopNodNormalisedModelRatiosTime_160.eps}
\caption{Photometric results of the measured calibration sources at 70, 100, and
160~$\mu$m (top to bottom) vs. time normalised to the mean flux ratio in each
band according to Tab.~\ref{tab:photstat}. The horizontal lines depict the mean
and the rms of the averaged photometric results for OD $\geq 300$ obtained for
the three wave bands. The red lines correspond to the fits according to
Eq.~\ref{eq:corrfit} and Tab.~\ref{tab:corrfit}.}
\label{fig:photnormfit}
\end{figure}

The measured flux, at 70~$\mu$m of the calibration objects
declines nearly linearly until OD~300. The same is visible for the other two
bands after normalising the individual sources to the same mean flux ratio
distribution (see Fig.~\ref{fig:photnormfit}). For this, the data of each of the
five objects have been divided by their individually derived mean flux ratios
measured in a given wave band and then multiplied with the mean of the complete
set of targets. It is interesting that such a behaviour is not seen for scan map
observations, but there were not many of such observations done during this
period, either. Furthermore, this drift is not detected when analysing the
signal of the internal calibration sources obtained during the calibration block
measurements \cite{moor13}. This points to an origin of the effect outside the
PACS instrument.

The telescope main mirror is the by far brightest background emitter
with an estimated mean flux of 200, 145 and 95 GJy/sr at 70, 100, and 160~$\mu$m
(Poglitsch, priv. comm.) that determines the average thermal load on the detectors. It
varies with time, mostly due to seasonal effects. The flux can be
modelled by using the measured mirror temperature as an input and including
certain assumptions for emissivity and its change during the Herschel mission.
Figure~\ref{fig:TM1} (upper panel) demonstrates how the mean temperature of the
telescope main mirror and its modelled flux at 70~$\mu$m varies with time.
We see that after the initial cool-down phase until OD~64, there is a
period of rapid warm-up until approximately OD~200.

The variation of the
total flux load on the detectors modifies their response. A brighter background
radiation leads to smaller measured point source fluxes. As
shown in \cite{balog13}, some of the scatter produced by the photometry of the
five standard stars observed with the scan map AOT can be mitigated, if the
evolution of the telescope background flux is considered. To investigate the
prospects of a similar correction for the chopped observations, we present
Fig.~\ref{fig:FvsFM1}, where the photometric results of $\gamma$~Dra are put into
perspective to the matching modelled telescope main mirror fluxes.
Unfortunately, the temporal coverage of the model does not extend below OD~150,
hence it misses most of the critical period, where the response deviated most.
Therefore, we cannot trace the entire parameter space to look for a supposed 
correlation.
Nonetheless, the plot demonstrates that it is separated into two
distinct populations. The intersection happens at a main mirror flux of around
190~GJy/sr at 70~$\mu$m, which coincides with the period around OD~300. It seems
that the short term variations and long term trend as of OD~300 of the chopped
photometer data may also be corrected for influences by the changing main mirror
flux, but not the strong decline in response at the beginning of the
Herschel mission as visualised in Fig.~\ref{fig:TM1} (lower panel). This effect
seems to be caused by yet another phenomenon.

\begin{figure}
\centering
\includegraphics[width=1.0\textwidth]{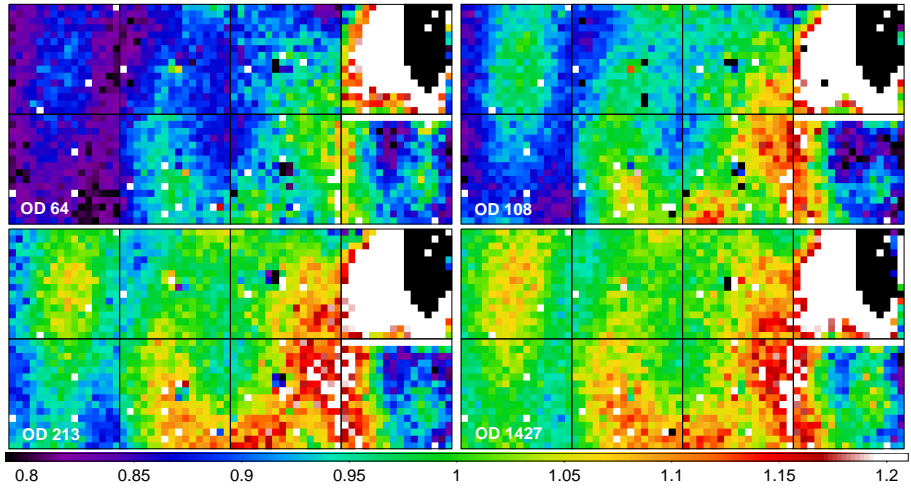}
\caption{
Visualisation of the differential signal variation of the short-wave
detector array at 70~$\mu$m caused by the changing telescope background
for four ODs 64, 108, 213, 1427. The eight individual sub-matrices are indicated
as black frames. The signals have been normalised to an observation obtained on
OD~1444. While a similar emission patterns can be recognised in all of the four
images, their scaling varies in time.}
\label{fig:M1emission}
\end{figure}

To find such influences, it is reasonable to look for causes that
are related to the observing mode. The main difference to scan maps is the
measurement at alternating ON and OFF positions during the chop-nod sequence.
Chopping is done with a focal plane mirror that modifies the viewing angle by
about $50''$ along the instrument y-axis.
To investigate a possible change in the differential signal that is produced
by the chopping technique, we have produced images for chop-nod observations
obtained at 70~$\mu$m, where the chopped OFF position was subtracted from the
chopped ON position. To be able to analyse the signal variation instead of the
absolute signal, we have divided each image by the one derived from the last
observation done on OD~1444 (see Eq.~\ref{eq:onoff}).

\begin{equation}
\frac{\left<\mathrm{ON}-\mathrm{OFF}\right>_i}{\left<\mathrm{ON}-\mathrm{OFF}
\right>_\mathrm{OD 1444}}
\label{eq:onoff}
\end{equation}

\begin{figure}
\centering
\psfrag{tod}[cc][][0.9][0]{$t$ [OD]}
\psfrag{F}[cc][][0.9][0]{$\Delta F(70)_\mathrm{back,norm}$}
\psfrag{od064}[][][0.9][0]{OD 64}
\psfrag{od200}[][][0.9][0]{OD 200}
\psfrag{od300}[][][0.9][0]{OD 300}
\includegraphics[width=0.9\textwidth]{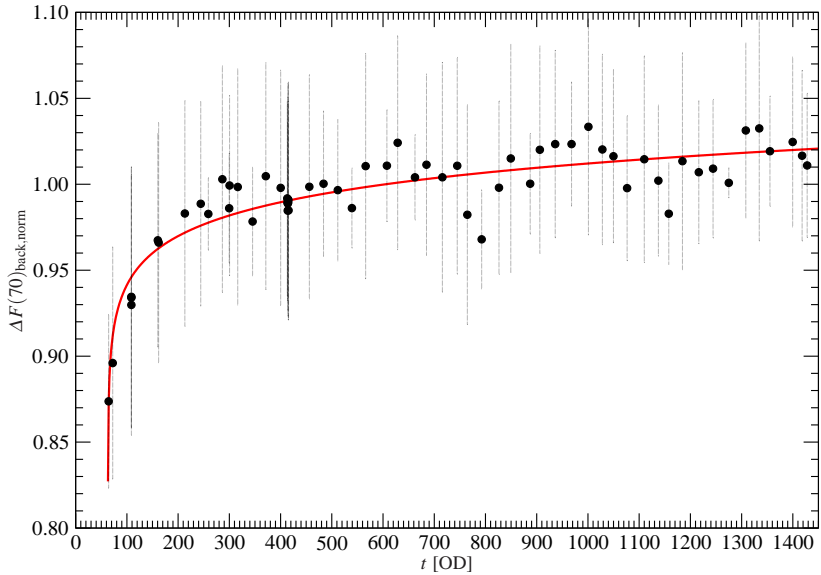}
\caption{Evolution of the mean of the chopped differential background signal at
70~$\mu$m normalised to the results of the OD~1444 observation and measured by
the two left detector matrices (see Fig.~\ref{fig:M1emission}). The error bars
represent the standard deviation of the signal distribution. The solid red line
is a fit to the data according to $\Delta F(70)_\mathrm{back,norm} = c_1 \cdot
(t-t_0)^{c_2}+c_3$.}
\label{fig:diffFM1}
\end{figure}

The results for four observations from ODs 64, 108, 213 and 1427 are shown in
Fig.~\ref{fig:M1emission}. They demonstrate a clear change of the differential
signal during the first phase of the Herschel mission. An even clearer picture
of the evolution of the differential signal is given by Fig.~\ref{fig:diffFM1}.
We have calculated the mean and the standard deviation of the normalised
differential signals of all chopped $\gamma$~Dra observations obtained at
70$\mu$m measured in the two left blue detector matrices (see
Figs.~\ref{fig:obsmode}, \ref{fig:M1emission}). This reflects the change of
the differential background flux seen by the PACS instrument.

Such a change in the differential signal can only be explained by
a stronger flux and temperature gradient between the two beams.
A temperature gradient across the telescope main mirror does not explain this
behaviour, because

\begin{enumerate}
\item the eight thermistors attached to it do not show any significant variation
in the pairwise differences of the measured temperature,
\item the areas on the mirror covered by the alternating beams during chopping
only deviate slightly. Even if the temperature of the main mirror had a strong
spatial gradient, the effect would be quite small.
\end{enumerate}

The strong temperature drift of the main mirror as shown in Fig.~\ref{fig:TM1}
(lower panel) and a similar behaviour of the secondary mirror suggest that the
whole telescope was not yet thermally stable, as it was after OD~300, so that
the alternating beam paths register a changing offset of the background flux.
This appears to be the root cause for the response drift of the detectors we
want to correct for.

This phenomenon is in so far quite striking, because it covers the PV
phase, in which the final tuning and initial calibration of the detectors were
supposed to happen. It seems that the conditions during that phase were
not quite comparable to those experienced during the Routine Science Phase 
(RSP).
Therefore, it is important to re-investigate the instrument calibration during
the Post-Operations Phase (POP), when all long term effects can be
analysed.

\section{Correction}
\label{sec:corr}

The discovered variation of the differential background flux suggests a similar
functional relationship of the response change with time instead of a simple and
physically unjustified piecewise linear evolution. For now, we neglect a direct 
correlation
between the response change and the variation of the background flux, mainly
because the current telescope model is not the final one. This may
be re-investigated in the near future. Instead, we empirically fit the data with
a relationship that is motivated by this measurement.

\begin{figure}
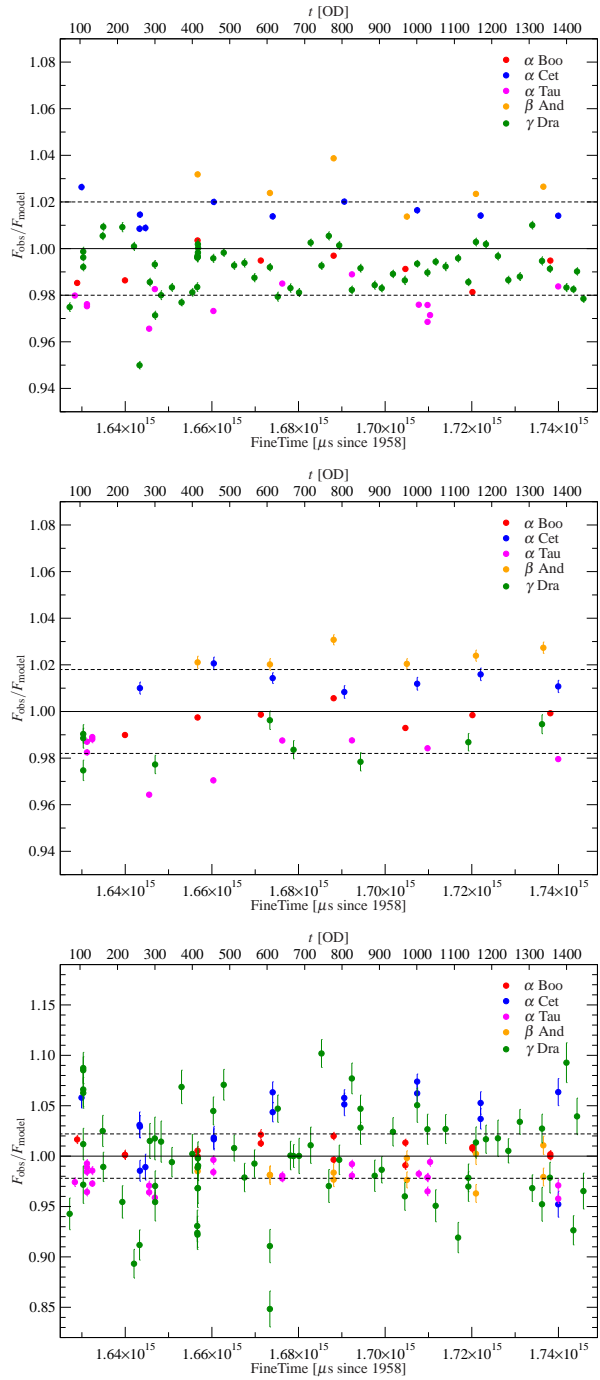

\centering
\psfrag{OD}[cc][][0.7][0]{$t$ [OD]}
\psfrag{FT}[cc][][0.7][0]{FineTime [$\mu$s since 1958]}
\psfrag{F}[cc][][0.7][0]{$F_\mathrm{obs}/F_\mathrm{model}$}
\psfrag{aBoo}[cc][][0.7][0]{$\alpha$~Boo}
\psfrag{aCet}[cc][][0.7][0]{$\alpha$~Cet}
\psfrag{aTau}[cc][][0.7][0]{$\alpha$~Tau}
\psfrag{bAnd}[cc][][0.7][0]{$\beta$~And}
\psfrag{gDra}[cc][][0.7][0]{$\gamma$~Dra}
\includegraphics[width=0.65\textwidth]{5FiducialChopNodScaledModelRatiosTime_70.eps}

\medskip
\includegraphics[width=0.65\textwidth]{5FiducialChopNodScaledModelRatiosTime_100.eps}

\medskip
\includegraphics[width=0.65\textwidth]{5FiducialChopNodScaledModelRatiosTime_160.eps}
\caption{Photometric results of the measured calibration sources at 70, 100,
and 160~$\mu$m (top to bottom) vs. time after applying the calibration 
correction.
The horizontal lines depict the mean and the rms derived from the averaged 
photometric
results of the individual objects.}
\label{fig:corrphotometry}
\end{figure}

Before being able to fit the photometric data, we first normalise them to their
combined mean flux ratios of each wave band according to 
Tab.~\ref{tab:photstat}. 
The flux ratios of each of the objects are first divided by their mean and then
multiplied by the global mean per wave band. In this way,
we are able to merge the five individual datasets per wavelength into a single
one, and at the same time, derive a result that is independent of the number of
observations per target (see Fig.~\ref{fig:photnormfit}).
We have removed the photometry of $\gamma$~Dra from
the fit at 160~$\mu$m because of the large scatter (see
Sect.~\ref{sec:results_late}). During the first attempts, we kept the exponent
of the power-law fit as a free parameter. However, it turned out to be close to
$-1$ in all cases, so we decided to fix it to this value and fit only two
parameters without affecting the reliability of the fit. Finally, the fits have
been derived according to

\begin{equation}
\frac{F_\mathrm{obs}}{F_\mathrm{model}} = c_1 + c_2 \left( 
\frac{\mathrm{FT}}{10^{15}}-1.62\right)^{-1}
\label{eq:corrfit}
\end{equation}

where FT is the time in \texttt{FineTime} units. The values of the fitting
parameters $c_1$ and $c_2$ are given in Tab.~\ref{tab:corrfit}.
The results are presented in Fig.~\ref{fig:photnormfit}. 

\begin{table}
\caption{Fit parameters according to Eq.~\ref{eq:corrfit}.}
\label{tab:corrfit}
\begin{tabular}{rccc}
\hline
\noalign{\smallskip}
Wavelength & \multicolumn{1}{c}{$c_1$} & \multicolumn{1}{c}{$c_2$} &
\multicolumn{1}{c}{Uncertainty}\\
\noalign{\smallskip}
\hline
\noalign{\smallskip}
 70 $\mu$m & $0.924$ & $6.551\times10^{-4}$ & 0.9\% \\
100 $\mu$m & $0.937$ & $4.276\times10^{-4}$ & 0.6\% \\
160 $\mu$m & $0.949$ & $1.948\times10^{-4}$ & 1.9\% \\
\noalign{\smallskip}
\hline
\end{tabular}
\end{table}

The results show that the correction is most important for the 70~$\mu$m band,
while at 160~$\mu$m it is only small, at least for the
objects we include in our calibration. For higher S/N data, the effect may be
more obvious. Table~\ref{tab:corrfit} together with Eq.~\ref{eq:corrfit} is used
to calculate the flux correction to the data. This is done by determining a
representative (the central) time of a given observation, express it in
\texttt{FineTime} units and compute Eq.~\ref{eq:corrfit} for the suitable
wavelength. For correcting the data, the signal values must be divided by the
result.

In order to verify the correction derived by the method described above, we 
have applied
it to the calibration observations. We modified the detector data directly by 
using
the task definition provided in Sect.~\ref{s:fluxcorr}. In order for it to work
properly, it must be placed after the pipeline tasks
\texttt{photRespFlatfieldCorrection} and \linebreak 
\texttt{photNonLinearityCorrection}.
However, the correction can also be done by applying it to the photometry 
results.

\begin{table}
\caption{Results of the photometry of the five prime standard stars after
applying the flux calibration correction. The quoted
numbers are the mean values of the flux ratios of all observations. When
combining more than one object, the mean is weighted by the number of objects.}
\label{tab:corrphotstat}
\begin{tabular}{lccc}
\hline\noalign{\smallskip}
& \multicolumn{3}{c}{$\left< F_\mathrm{obs}/F_\mathrm{model} \right>$} \\
\multicolumn{1}{c}{Object(s)} & \multicolumn{1}{c}{70 $\mu$m} & 
\multicolumn{1}{c}{100 $\mu$m} & \multicolumn{1}{c}{160 $\mu$m} \\
\noalign{\smallskip}\hline\noalign{\smallskip}
$\alpha$~Boo & $0.992 \pm 0.007$ & $0.997 \pm 0.005$ & $1.006 \pm 0.009$ \\ 
$\alpha$~Cet & $1.016 \pm 0.005$ & $1.013 \pm 0.004$ & $1.034 \pm 0.033$ \\
$\alpha$~Tau & $0.977 \pm 0.007$ & $0.982 \pm 0.008$ & $0.978 \pm 0.011$ \\
$\beta$~And  & $1.026 \pm 0.008$ & $1.024 \pm 0.004$ & $0.986 \pm 0.013$ \\
$\gamma$~Dra & $0.991 \pm 0.010$ & $0.986 \pm 0.008$ & $0.997 \pm 0.051$ \\
\noalign{\smallskip}\hline\noalign{\smallskip}
K giants     & $0.987 \pm 0.008$ & $0.988 \pm 0.008$ & $0.994 \pm 0.014$ \\
M giants     & $1.021 \pm 0.008$ & $1.019 \pm 0.008$ & $1.010 \pm 0.035$ \\
\noalign{\smallskip}\hline\noalign{\smallskip}
all          & $1.000 \pm 0.020$ & $1.000 \pm 0.018$ & $1.000 \pm 0.022$ \\
\noalign{\smallskip}\hline
\end{tabular}
\end{table}

The new photometric results after applying the flux correction are shown in 
Fig.~
\ref{fig:corrphotometry}. The statistics are summarised in
Tab.~\ref{tab:corrphotstat}. 
{
In addition, the corrected flux values of all observations
are listed in Tab.~\ref{tab:photlist}.
}
The main results are that

\begin{itemize}
	\item the mean ratio between the model and the measured flux is 1 when
		weighting the data according to the mean of the individual
		targets,
	\item the initial response change has disappeared.
\end{itemize}

One might get the impression that the data have been slightly over-corrected at
100~$\mu$m, but this is caused by the low number statistics, the lower S/N of
the fainter objects and the spread among the individual objects. Note that the
horizontal lines in the plots representing the mean and rms 
are derived from the mean values calculated for a given target. The rms does not
reflect the reproducibility of the photometry, which itself is of the order of
1\%, if the S/N is high enough (see Tab.~\ref{tab:corrphotstat}).

\section{Discussion and Caveats}
\label{s:caveats}

\subsection{Model fluxes}
\label{s:modflux}

The stellar spectra have been modelled by assuming for each star a constant
stellar radius for all wavelengths. A correct modelling would have to include
optical depth effects that are wavelength dependent and result in varying
effective stellar radii. Correcting for this effect would introduce flux changes
of the order of 2\%. This will be included in future flux calibrations.

\subsection{Empirical approach}
\label{s:empirical}

Although the flux correction is motivated by a physical change of the measured
differential background flux
that is most likely responsible for the response drift at the beginning of the
Herschel mission, the correction itself is a pure empirical approach with some
ad hoc assumptions. In particular, the origin of the scaling factor between
chop-nod and scan map observations beyond OD~300 is not fully understood,
although it is believed that to some extent the different signal modulation
frequencies play a role.
Nevertheless, we are certain that the flux correction as derived here is
reliable and at least reduces the systematic uncertainties due to the mismatch
introduced by the scan-map-based flux calibration. The PACS Instrument Control
Centre (ICC) is working on
deriving a realistic time-dependent flux model of the telescope main mirror that
may help to re-iterate on this issue in the future.

\subsection{Aperture correction}
\label{s:psf}

The aperture correction applied to the photometry is based on an encircled
energy function of a template PSF that was derived from scan map observations.
We already know that the PSF derived from scan map observations is not even
universal for all types of scan maps, let alone can it be regarded as a good
representation of the PSF produced by chop-nod observations. However, the
differences are small and mostly affect the wings. Therefore, by
selecting a large aperture, the uncertainties introduced by the aperture
correction are minimised. The remaining systematic differences are
contained in the flux correction factors.

Theoretically, one would have to derive proper PSFs from point source
observations using the chopped photometry AOT. However, this is compromised
by a small effective field-of-view (FOV) in the maps, so that
the PSFs of the neighbouring images (see Fig.~\ref{fig:map}) may prevent 
establishing a suitable
template PSF from the central image (see also Sect.~\ref{s:fluxdep}).

\subsection{Flux dependency}
\label{s:fluxdep}

At the moment, the flux correction for chop-nod observations has only been
verified for a restricted range of source fluxes, i.~e.~the prime calibrators.
Further systematic verifications using much brighter objects (e.~g.~planets, 
asteroids) is
still an outstanding issue. First tests seem to confirm that such a flux
dependency might be negligible \cite{mueller13}.

One possible source of introducing such a suggested flux dependency is the 
limited FOV of
the point-source maps. The patches in the map belonging to the individual
positive and negative images of the object are smaller than their measurable 
PSF. This
potentially can lead to a contamination of the background level of the central
patch that is used for the photometry. Such a contamination may depend on the
brightness of a given object. For rather faint objects, the PSF declines quickly
enough so that contributions from the surrounding images to the central image
are well below the noise level. This is not the case for very bright objects or
high S/N data. The potential amount of contamination is yet to be established.

\section{Conclusions}
\label{s:conclusions}

We have presented a reliable flux calibration for the chopped point-source
photometry AOT observing mode of the PACS photometer on-board the Herschel Space
Observatory, whose flux calibration is tuned to the preferred observing mode,
i.~e.~scan maps.  This was done by scaling the photometric results of the
five PACS prime standard stars $\alpha$~Boo, $\alpha$~Cet, $\alpha$~Tau,
$\beta$~And, and $\gamma$~Dra obtained from chopped observations to the scan-map
based calibration scheme.

We found a
strong decline of the measured integrated point-source flux densities for the
first 300 Operational Days that is not seen for scan map observations. If only
the results beyond that threshold are considered, the photometry produces
results that are on average lower by $5-6$\% than what is obtained from scan map
observations. In addition, we find that the K and M giants in the sample
produce slightly different results of the order $2-3$\% in all bands.

In an attempt to identify the reason for the initial rapid change of the
effective signal response, we have found an equally strong variation of the
measured differential background flux that indicates a systematic change of the
temperature or flux gradient seen by the alternating instrument beam. This
variation seems to be anti-correlated with the photometric results. It is known
that a higher background flux reduces the response of the bolometers. The root
cause that is responsible for changing the gradient in the background flux is
unknown, but we were able to restrict it to an origin outside the PACS
instrument.

After fitting an empirical relationship between the measured ratios of
the model and photometric fluxes for every calibration object and in each band,
the initial response drift has disappeared. In addition, the residual scatter
of the modelled-to-\linebreak measured flux ratios of the five calibrators are 
consistent
within 2\%, which is well below the quoted model uncertainty of 5\%. The
reproducibility for a given object is of the order 1\% at 70 and 100~$\mu$m and
varies between 1\% and 5\% at 160~$\mu$m. The latter can be explained by a
decreasing signal-to-noise ratio.

This demonstrates the reliability of the established flux calibration that is of
the same order as achieved for the scan map observing mode.

\begin{acknowledgements}
We thank the anonymous referee for the constructive report on the submitted
manuscript. We also would like to thank J. Blommaert (KUL) for providing helpful
comments. Z. Balog, H. Linz and M. Nielbock are funded by the Deutsches Zentrum
f\"ur Luft- und Raumfahrt e. V.
\end{acknowledgements}

\bibliographystyle{spmpsci}
\bibliography{references}

\clearpage
\appendix
\section{List of Observations}
\label{sec:obslist}

\begin{table}[!h]
\caption{List of observations. The repetition factor was always set to 1. The
first four observations and the one with OBSID 1342184286 were carried out with
low gain settings. This had no influence on the photometry.}
\label{tab:obslist}
\begin{tabular}{rll}
\hline \noalign{\smallskip}
\multicolumn{1}{c}{OD} & \multicolumn{1}{c}{OBSID} & \multicolumn{1}{c}{AOR 
label} \\
\noalign{\smallskip} \hline \noalign{\smallskip}
72 & 1342180683 & PVPhotFPG\_261C\_StdPS\_blu\_SAA-20-10\_HIP87833\\
86 & 1342181662 & PVPhotFPG\_261F\_StdPS\_blu\_SAA+20+30\_HIP21421\_0001\\
92 & 1342182100 & PVPhotFPG\_261G\_StdPS\_blu\_SAA\_0+30\_HIP69673\_OD92\_01\\
104 & 1342182830 & PVPhotFPG\_261C\_PS\_blu\_SAA-20-10\_HIP14135\_OD104\_01\\
108 & 1342182990 & PVPhotAOTVal\_511A\_StdPSdith\_blu\_gammaDra\_0001\\
108 & 1342182991 & PVPhotAOTVal\_511A\_StdPSndith\_blu\_gammaDra\_0001\\
108 & 1342182993 & PVPhotFlux\_321C\_StdPS\_photcal\_blu\_gammaDra\_0001\\
108 & 1342182994 & PVPhotAOTVal\_511A\_StdPSndith\_grn\_gammaDra\_0001\\
108 & 1342182995 & PVPhotFlux\_321C\_StdPS\_photcal\_grn\_gammaDra\_0001\\
108 & 1342182996 & PVPhotAOTVal\_511A\_StdPSdith\_grn\_gammaDra\_0001\\
118 & 1342183530 & PVPhotFlux\_321C\_StdPS\_photcal\_blu\_alfTau\_0001\\
118 & 1342183531 & PVPhotFlux\_321C\_StdPS\_photcal\_grn\_alfTau\_0001\\
118 & 1342183536 & PVPhotSpatial\_314B\_StdPSdith\_blu\_AlfTau\_0001\\
118 & 1342183537 & PVPhotSpatial\_314B\_StdPSdith\_grn\_AlfTau\_0001\\
132 & 1342184285 & PVPhotFlux\_324A\_StdPS\_hi10Jy\_grn\_alfTau\_0002\\
132 & 1342184286 & PVPhotFlux\_323A\_StdPS\_lo10Jy\_grn\_alfTau\_0002\\
160 & 1342186140 & PVPhotFlux\_321B\_StdPS\_photcal\_blu\_gamDra\_0001\\
161 & 1342186191 & PVPhotFlux\_321B\_StdPS\_photcal\_blu\_gamDra\_0002\\
213 & 1342188069 & RPPhotFlux\_321A\_cPS\_repro\_blu\_gamDra\_0001\\
220 & 1342188243 & RPPhotFlux\_321C\_cPS\_photcal\_blu\_alfBoo\_0001\\
220 & 1342188244 & RPPhotFlux\_321C\_cPS\_photcal\_grn\_alfBoo\_0001\\
244 & 1342189188 & RPPhotFlux\_321A\_cPS\_repro\_blu\_gamDra\_0002\\
258 & 1342189775 & RPPhotFPG\_261A\_StdPS\_blu\_SAA-20+30\_OD258\_HIP87833\\
258 & 1342189783 & RPPhotFPG\_261A\_StdPS\_blu\_SAA-20+30\_OD258\_HIP14135\\
259 & 1342189823 & RPPhotFlux\_321C\_cPS\_photcal\_blu\_alfCet\_0001\\
259 & 1342189826 & RPPhotFlux\_321C\_cPS\_photcal\_grn\_alfCet\_0001\\
274 & 1342190603 & RPPhotFPG\_261A\_cPS\_blu\_SAA-20+30\_OD274\_ni\_HIP14135\\
284 & 1342190943 & RPPhotFlux\_324A\_cPS\_10Jy\_grn\_alfTau\_0001\\
284 & 1342190946 & RPPhotFlux\_324A\_cPS\_20Jy\_blu\_alfTau\_0001\\
286 & 1342191124 & RPPhotFlux\_321A\_cPS\_repro\_blu\_gamDra\_0003\\
299 & 1342191848 & RPPhotFPG\_261B\_cPS\_blu\_SAA-20+30\_OD299\_ni\_HIP87833\\
299 & 1342191870 & RPPhotFPG\_261B\_cPS\_blu\_SAA-20+30\_OD299\_ni\_HIP21421\\
300 & 1342191957 & RPPhotFlux\_321A\_cPS\_repro\_blu\_gamDra\_0004\\
300 & 1342191960 & RPPhotFlux\_324A\_cPS\_2Jy\_grn\_gamDra\_0001\\
316 & 1342192779 & RPPhotFlux\_321A\_cPS\_repro\_blu\_gamDra\_0005\\
345 & 1342195482 & RPPhotFlux\_321A\_cPS\_repro\_blu\_gamDra\_0006\\
371 & 1342196729 & RPPhotFlux\_321A\_cPS\_repro\_blu\_gamDra\_0007\\
400 & 1342198498 & RPPhotFlux\_321A\_cPS\_repro\_blu\_gamDra\_0008\\
413 & 1342199480 & RPPhotFlux\_321A\_cPS\_repro\_blu\_gamDra\_0009\\
413 & 1342199511 & RPPhotFlux\_321A\_cPS\_repro\_blu\_gamDra\_0010\\
413 & 1342199525 & RPPhotFlux\_321A\_cPS\_repro\_blu\_gamDra\_0011\\
414 & 1342199599 & RPPhotFlux\_321A\_cPS\_repro\_blu\_gamDra\_0012\\
414 & 1342199602 & RPPhotFlux\_324A\_cPS\_20Jy\_blu\_alfBoo\_0001\\
414 & 1342199605 & RPPhotFlux\_324A\_cPS\_10Jy\_grn\_alfBoo\_0001\\
414 & 1342199608 & RPPhotFlux\_321C\_cPS\_photcal\_blu\_betAnd\_0001\\
414 & 1342199611 & RPPhotFlux\_321C\_cPS\_photcal\_grn\_betAnd\_0001\\
414 & 1342199638 & RPPhotFlux\_321A\_cPS\_repro\_blu\_gamDra\_0013\\
\noalign{\smallskip}\hline
\end{tabular}
\end{table}

\addtocounter{table}{-1}
\begin{table}
\caption{continued.}
\begin{tabular}{rll}
\hline \noalign{\smallskip}
\multicolumn{1}{c}{OD} & \multicolumn{1}{c}{OBSID} & \multicolumn{1}{c}{AOR 
label} \\
\noalign{\smallskip} \hline \noalign{\smallskip}
414 & 1342199654 & RPPhotFlux\_321A\_cPS\_repro\_blu\_gamDra\_0014\\
415 & 1342199706 & RPPhotFlux\_321A\_cPS\_repro\_blu\_gamDra\_0015\\
415 & 1342199716 & RPPhotFlux\_321A\_cPS\_repro\_blu\_gamDra\_0016\\
456 & 1342202941 & RPPhotFlux\_321A\_cPS\_repro\_blu\_gamDra\_0018\\
456 & 1342202957 & RPPhotFlux\_324A\_cPS\_10Jy\_grn\_alfTau\_0002\\
456 & 1342202960 & RPPhotFlux\_324A\_cPS\_20Jy\_blu\_alfTau\_0002\\
457 & 1342203029 & RPPhotFlux\_324A\_cPS\_5Jy\_blu\_alfCet\_0001\\
457 & 1342203032 & RPPhotFlux\_324A\_cPS\_2Jy\_grn\_alfCet\_0001\\
483 & 1342204208 & RPPhotFlux\_321A\_cPS\_repro\_blu\_gamDra\_0019\\
511 & 1342206000 & RPPhotFlux\_321A\_cPS\_repro\_blu\_gamDra\_0020\\
539 & 1342208970 & RPPhotFlux\_321A\_cPS\_repro\_blu\_gamDra\_0021\\
566 & 1342210581 & RPPhotFlux\_321A\_cPS\_repro\_blu\_gamDra\_0022\\
583 & 1342211279 & RPPhotFlux\_324A\_cPS\_20Jy\_blu\_alfBoo\_0002\\
583 & 1342211282 & RPPhotFlux\_324A\_cPS\_10Jy\_grn\_alfBoo\_0002\\
607 & 1342212493 & RPPhotFlux\_321A\_cPS\_repro\_blu\_gamDra\_0023\\
607 & 1342212496 & RPPhotFlux\_324A\_cPS\_2Jy\_grn\_gamDra\_0002\\
607 & 1342212503 & RPPhotFlux\_324A\_cPS\_2Jy\_grn\_betaAnd\_0001\\
607 & 1342212506 & RPPhotFlux\_324A\_cPS\_5Jy\_blu\_betaAnd\_0001\\
614 & 1342212852 & RPPhotFlux\_324A\_cPS\_2Jy\_grn\_alfCet\_0002\\
614 & 1342212855 & RPPhotFlux\_324A\_cPS\_5Jy\_blu\_alfCet\_0002\\
628 & 1342213587 & RPPhotFlux\_321A\_cPS\_repro\_blu\_gamDra\_0024\\
640 & 1342214210 & RPPhotFlux\_324A\_cPS\_20Jy\_blu\_alfTau\_0003\\
640 & 1342214213 & RPPhotFlux\_324A\_cPS\_10Jy\_grn\_alfTau\_0003\\
662 & 1342215373 & RPPhotFlux\_321A\_cPS\_repro\_blu\_gamDra\_0025\\
670 & 1342216068 & RPPhotFlux\_324A\_cPS\_2Jy\_grn\_gamDra\_0003\\
684 & 1342217403 & RPPhotFlux\_321A\_cPS\_repro\_blu\_gamDra\_0026\\
715 & 1342220822 & RPPhotFlux\_321A\_cPS\_repro\_blu\_gamDra\_0027\\
744 & 1342221810 & RPPhotFlux\_321A\_cPS\_repro\_blu\_gamDra\_0028\\
764 & 1342222755 & RPPhotFlux\_321A\_cPS\_repro\_blu\_gamDra\_0029\\
777 & 1342223334 & RPPhotFlux\_324A\_cPS\_5Jy\_blu\_betaAnd\_0003\\
777 & 1342223337 & RPPhotFlux\_324A\_cPS\_2Jy\_grn\_betaAnd\_0003\\
777 & 1342223344 & RPPhotFlux\_324A\_cPS\_20Jy\_blu\_alfBoo\_0003\\
777 & 1342223347 & RPPhotFlux\_324A\_cPS\_10Jy\_grn\_alfBoo\_0003\\
792 & 1342224228 & RPPhotFlux\_321A\_cPS\_repro\_blu\_gamDra\_0030\\
806 & 1342224926 & RPPhotFlux\_324A\_cPS\_5Jy\_blu\_alfCet\_0003\\
806 & 1342224929 & RPPhotFlux\_324A\_cPS\_2Jy\_grn\_alfCet\_0003\\
826 & 1342226711 & RPPhotFlux\_321A\_cPS\_repro\_blu\_gamDra\_0031\\
826 & 1342226739 & RPPhotFlux\_324A\_cPS\_20Jy\_blu\_alfTau\_0004\\
826 & 1342226742 & RPPhotFlux\_324A\_cPS\_10Jy\_grn\_alfTau\_0004\\
849 & 1342228387 & RPPhotFlux\_321A\_cPS\_repro\_blu\_gamDra\_0032\\
849 & 1342228390 & RPPhotFlux\_324A\_cPS\_2Jy\_grn\_gamDra\_0004\\
887 & 1342231096 & RPPhotFlux\_321A\_cPS\_repro\_blu\_gamDra\_0033\\
906 & 1342231898 & RPPhotFlux\_321A\_cPS\_repro\_blu\_gamDra\_0034\\
936 & 1342234213 & RPPhotFlux\_321A\_cPS\_repro\_blu\_gamDra\_0035\\
967 & 1342237974 & RPPhotFlux\_321A\_cPS\_repro\_blu\_gamDra\_0036\\
969 & 1342236964 & RPPhotFlux\_324A\_cPS\_10Jy\_grn\_alfBoo\_0004\\
969 & 1342236967 & RPPhotFlux\_324A\_cPS\_20Jy\_blu\_alfBoo\_0004\\
973 & 1342237160 & RPPhotFlux\_324A\_cPS\_2Jy\_grn\_betaAnd\_0004\\
973 & 1342237163 & RPPhotFlux\_324A\_cPS\_5Jy\_blu\_betaAnd\_0004\\
1000 & 1342238771 & RPPhotFlux\_321A\_cPS\_repro\_blu\_gamDra\_0037\\
1000 & 1342238778 & RPPhotFlux\_324A\_cPS\_2Jy\_grn\_alfCet\_0004\\
1000 & 1342238781 & RPPhotFlux\_324A\_cPS\_5Jy\_blu\_alfCet\_0004\\
1005 & 1342239042 & ObsCal\_RP\_FPG\_PPhot\_Blue\_cycle59\_OD1005\_HIP21421\\
1028 & 1342240698 & RPPhotFlux\_321A\_cPS\_repro\_blu\_gamDra\_0038\\
\noalign{\smallskip}\hline
\end{tabular}
\end{table}

\addtocounter{table}{-1}
\begin{table}
\caption{continued.}
\begin{tabular}{rll}
\hline \noalign{\smallskip}
\multicolumn{1}{c}{OD} & \multicolumn{1}{c}{OBSID} & \multicolumn{1}{c}{AOR 
label} \\
\noalign{\smallskip} \hline \noalign{\smallskip}
1028 & 1342240753 & ObsCal\_RP\_FPG\_PPhot\_Blue\_cycle60\_OD1028\_HIP21421\\
1028 & 1342240754 & RPPhotFlux\_324A\_cPS\_20Jy\_blu\_alfTau\_0005\\
1028 & 1342240757 & RPPhotFlux\_324A\_cPS\_10Jy\_grn\_alfTau\_0005\\
1034 & 1342241329 & ObsCal\_RP\_FPG\_PPhot\_Blue\_cycle61\_OD1034\_HIP21421\\
1049 & 1342242556 & RPPhotFlux\_321A\_cPS\_repro\_blu\_gamDra\_0039\\
1076 & 1342244899 & RPPhotFlux\_321A\_cPS\_repro\_blu\_gamDra\_0040\\
1109 & 1342246180 & RPPhotFlux\_321A\_cPS\_repro\_blu\_gamDra\_0041\\
1137 & 1342247334 & RPPhotFlux\_321A\_cPS\_repro\_blu\_gamDra\_0042\\
1137 & 1342247337 & RPPhotFlux\_324A\_cPS\_2Jy\_grn\_gamDra\_0005\\
1148 & 1342247701 & RPPhotFlux\_324A\_cPS\_10Jy\_grn\_alfBoo\_0005\\
1148 & 1342247704 & RPPhotFlux\_324A\_cPS\_20Jy\_blu\_alfBoo\_0005\\
1157 & 1342248031 & RPPhotFlux\_324A\_cPS\_2Jy\_grn\_betaAnd\_0005\\
1157 & 1342248034 & RPPhotFlux\_324A\_cPS\_5Jy\_blu\_betaAnd\_0005\\
1157 & 1342248037 & RPPhotFlux\_321A\_cPS\_repro\_blu\_gamDra\_0043\\
1170 & 1342248718 & RPPhotFlux\_324A\_cPS\_2Jy\_grn\_alfCet\_0005\\
1170 & 1342248721 & RPPhotFlux\_324A\_cPS\_5Jy\_blu\_alfCet\_0005\\
1184 & 1342249292 & RPPhotFlux\_321A\_cPS\_repro\_blu\_gamDra\_0044\\
1216 & 1342250855 & RPPhotFlux\_321A\_cPS\_repro\_blu\_gamDra\_0045\\
1244 & 1342252804 & RPPhotFlux\_321A\_cPS\_repro\_blu\_gamDra\_0046\\
1275 & 1342254722 & RPPhotFlux\_321A\_cPS\_repro\_blu\_gamDra\_0047\\
1308 & 1342256958 & RPPhotFlux\_321A\_cPS\_repro\_blu\_gamDra\_0048\\
1334 & 1342258830 & RPPhotFlux\_321A\_cPS\_repro\_blu\_gamDra\_0049\\
1334 & 1342258833 & RPPhotFlux\_324A\_cPS\_2Jy\_grn\_gamDra\_0006\\
1337 & 1342259256 & RPPhotFlux\_324A\_cPS\_2Jy\_grn\_betaAnd\_0006\\
1337 & 1342259259 & RPPhotFlux\_324A\_cPS\_5Jy\_blu\_betaAnd\_0006\\
1355 & 1342262224 & RPPhotFlux\_321A\_cPS\_repro\_blu\_gamDra\_0050\\
1356 & 1342262515 & RPPhotFlux\_324A\_cPS\_10Jy\_grn\_alfBoo\_0006\\
1356 & 1342262518 & RPPhotFlux\_324A\_cPS\_20Jy\_blu\_alfBoo\_0006\\
1377 & 1342263906 & RPPhotFlux\_324A\_cPS\_2Jy\_grn\_alfCet\_0006\\
1377 & 1342263909 & RPPhotFlux\_324A\_cPS\_5Jy\_blu\_alfCet\_0006\\
1377 & 1342263914 & RPPhotFlux\_324A\_cPS\_10Jy\_grn\_alfTau\_0006\\
1377 & 1342263917 & RPPhotFlux\_324A\_cPS\_20Jy\_blu\_alfTau\_0006\\
1399 & 1342267290 & RPPhotFlux\_321A\_cPS\_repro\_blu\_gamDra\_0051\\
1418 & 1342268965 & RPPhotFlux\_321A\_cPS\_repro\_blu\_gamDra\_0052\\
1427 & 1342269811 & RPPhotFlux\_321A\_cPS\_repro\_blu\_gamDra\_0053\\
1444 & 1342270999 & RPPhotFlux\_321A\_cPS\_repro\_blu\_gamDra\_0054\\
\noalign{\smallskip}\hline
\end{tabular}
\end{table}

\clearpage
\section{Photometry of Observations}
\label{sec:photlist}

\begin{table}[!h]
\caption{Photometry of all calibration observations.}
\label{tab:photlist}
\begin{tabular}{ll@{\hspace{1cm}}cc@{\hspace{1cm}}cc@{\hspace{1cm}}cc}
\hline \noalign{\smallskip}
 & & \multicolumn{6}{c}{Measured flux with correction applied [Jy]} \\
\multicolumn{1}{c}{OBSID} & \multicolumn{1}{c}{Object} & 
$F_{70}$ & $\sigma_{70}$ &
$F_{100}$ & $\sigma_{100}$ &
$F_{160}$ & $\sigma_{160}$ \\
\noalign{\smallskip} \hline \noalign{\smallskip}
1342182100 & $\alpha$~Boo & 15.207 & 0.047 &  &  & 2.939 & 0.092\\
1342188243 & $\alpha$~Boo & 15.224 & 0.036 &  &  & 2.895 & 0.104\\
1342188244 & $\alpha$~Boo &  &  & 7.433 & 0.033 & 2.894 & 0.080\\
1342199602 & $\alpha$~Boo & 15.488 & 0.045 &  &  & 2.907 & 0.093\\
1342199605 & $\alpha$~Boo &  &  & 7.490 & 0.031 & 2.891 & 0.083\\
1342211279 & $\alpha$~Boo & 15.355 & 0.049 &  &  & 2.927 & 0.069\\
1342211282 & $\alpha$~Boo &  &  & 7.499 & 0.032 & 2.953 & 0.111\\
1342223344 & $\alpha$~Boo & 15.387 & 0.052 &  &  & 2.881 & 0.065\\
1342223347 & $\alpha$~Boo &  &  & 7.552 & 0.027 & 2.948 & 0.085\\
1342236964 & $\alpha$~Boo &  &  & 7.456 & 0.032 & 2.929 & 0.080\\
1342236967 & $\alpha$~Boo & 15.299 & 0.044 &  &  & 2.865 & 0.094\\
1342247701 & $\alpha$~Boo &  &  & 7.497 & 0.031 & 2.916 & 0.065\\
1342247704 & $\alpha$~Boo & 15.146 & 0.046 &  &  & 2.912 & 0.090\\
1342262515 & $\alpha$~Boo &  &  & 7.503 & 0.030 & 2.890 & 0.075\\
1342262518 & $\alpha$~Boo & 15.354 & 0.046 &  &  & 2.896 & 0.084\\
\noalign{\smallskip} \hline \noalign{\smallskip}
1342182830 & $\alpha$~Cet & 5.018 & 0.039 &  &  & 0.982 & 0.075\\
1342189783 & $\alpha$~Cet & 4.931 & 0.039 &  &  & 0.957 & 0.093\\
1342189823 & $\alpha$~Cet & 4.960 & 0.035 &  &  & 0.915 & 0.075\\
1342189826 & $\alpha$~Cet &  &  & 2.417 & 0.026 & 0.955 & 0.079\\
1342190603 & $\alpha$~Cet & 4.932 & 0.041 &  &  & 0.918 & 0.093\\
1342203029 & $\alpha$~Cet & 4.987 & 0.036 &  &  & 0.943 & 0.076\\
1342203032 & $\alpha$~Cet &  &  & 2.442 & 0.027 & 0.945 & 0.079\\
1342212852 & $\alpha$~Cet &  &  & 2.427 & 0.023 & 0.968 & 0.070\\
1342212855 & $\alpha$~Cet & 4.957 & 0.036 &  &  & 0.987 & 0.076\\
1342224926 & $\alpha$~Cet & 4.988 & 0.034 &  &  & 0.982 & 0.059\\
1342224929 & $\alpha$~Cet &  &  & 2.413 & 0.028 & 0.976 & 0.081\\
1342238778 & $\alpha$~Cet &  &  & 2.421 & 0.028 & 0.997 & 0.054\\
1342238781 & $\alpha$~Cet & 4.969 & 0.038 &  &  & 0.986 & 0.076\\
1342248718 & $\alpha$~Cet &  &  & 2.431 & 0.027 & 0.962 & 0.073\\
1342248721 & $\alpha$~Cet & 4.958 & 0.036 &  &  & 0.977 & 0.081\\
1342263906 & $\alpha$~Cet &  &  & 2.419 & 0.026 & 0.884 & 0.093\\
1342263909 & $\alpha$~Cet & 4.958 & 0.035 &  &  & 0.987 & 0.097\\
\noalign{\smallskip} \hline \noalign{\smallskip}
1342181662 & $\alpha$~Tau & 13.846 & 0.041 &  &  & 2.608 & 0.090\\
1342183530 & $\alpha$~Tau & 13.794 & 0.044 &  &  & 2.657 & 0.081\\
1342183531 & $\alpha$~Tau &  &  & 6.788 & 0.033 & 2.647 & 0.076\\
1342183536 & $\alpha$~Tau & 13.782 & 0.038 &  &  & 2.636 & 0.079\\
1342183537 & $\alpha$~Tau &  &  & 6.819 & 0.029 & 2.582 & 0.083\\
1342184285 & $\alpha$~Tau &  &  & 6.833 & 0.033 & 2.604 & 0.060\\
1342184286 & $\alpha$~Tau &  &  & 6.827 & 0.044 & 2.639 & 0.084\\
1342190943 & $\alpha$~Tau &  &  & 6.662 & 0.028 & 2.599 & 0.095\\
1342190946 & $\alpha$~Tau & 13.645 & 0.040 &  &  & 2.581 & 0.079\\
1342191870 & $\alpha$~Tau & 13.885 & 0.047 &  &  & 2.567 & 0.103\\
1342202957 & $\alpha$~Tau &  &  & 6.705 & 0.031 & 2.634 & 0.078\\
1342202960 & $\alpha$~Tau & 13.753 & 0.048 &  &  & 2.667 & 0.090\\
1342214210 & $\alpha$~Tau & 13.919 & 0.053 &  &  & 2.618 & 0.071\\
1342214213 & $\alpha$~Tau &  &  & 6.823 & 0.028 & 2.625 & 0.080\\
\noalign{\smallskip}\hline
\end{tabular}
\end{table}

\addtocounter{table}{-1}
\begin{table}[!h]
\caption{continued.}
\begin{tabular}{ll@{\hspace{1cm}}cc@{\hspace{1cm}}cc@{\hspace{1cm}}cc}
\hline \noalign{\smallskip}
 & & \multicolumn{6}{c}{Measured flux with correction applied [Jy]} \\
\multicolumn{1}{c}{OBSID} & \multicolumn{1}{c}{Object} & 
$F_{70}$ & $\sigma_{70}$ &
$F_{100}$ & $\sigma_{100}$ &
$F_{160}$ & $\sigma_{160}$ \\
\noalign{\smallskip} \hline \noalign{\smallskip}
1342226739 & $\alpha$~Tau & 13.975 & 0.049 &  &  & 2.656 & 0.082\\
1342226742 & $\alpha$~Tau &  &  & 6.823 & 0.027 & 2.625 & 0.081\\
1342239042 & $\alpha$~Tau & 13.791 & 0.048 &  &  & 2.630 & 0.076\\
1342240753 & $\alpha$~Tau & 13.687 & 0.040 &  &  & 2.620 & 0.074\\
1342240754 & $\alpha$~Tau & 13.789 & 0.049 &  &  & 2.620 & 0.087\\
1342240757 & $\alpha$~Tau &  &  & 6.800 & 0.029 & 2.583 & 0.092\\
1342241329 & $\alpha$~Tau & 13.728 & 0.047 &  &  & 2.661 & 0.088\\
1342263914 & $\alpha$~Tau &  &  & 6.768 & 0.027 & 2.564 & 0.101\\
1342263917 & $\alpha$~Tau & 13.901 & 0.045 &  &  & 2.599 & 0.116\\
\noalign{\smallskip} \hline \noalign{\smallskip}
1342199608 & $\beta$~And & 5.772 & 0.035 &  &  & 1.050 & 0.069\\
1342199611 & $\beta$~And &  &  & 2.795 & 0.030 & 1.046 & 0.076\\
1342212503 & $\beta$~And &  &  & 2.792 & 0.029 & 1.042 & 0.077\\
1342212506 & $\beta$~And & 5.727 & 0.040 &  &  & 1.042 & 0.068\\
1342223334 & $\beta$~And & 5.811 & 0.036 &  &  & 1.037 & 0.058\\
1342223337 & $\beta$~And &  &  & 2.821 & 0.025 & 1.045 & 0.075\\
1342237160 & $\beta$~And &  &  & 2.793 & 0.025 & 1.060 & 0.059\\
1342237163 & $\beta$~And & 5.671 & 0.037 &  &  & 1.037 & 0.066\\
1342248031 & $\beta$~And &  &  & 2.803 & 0.027 & 1.023 & 0.074\\
1342248034 & $\beta$~And & 5.725 & 0.037 &  &  & 1.064 & 0.089\\
1342259256 & $\beta$~And &  &  & 2.812 & 0.028 & 1.040 & 0.074\\
1342259259 & $\beta$~And & 5.743 & 0.037 &  &  & 1.073 & 0.075\\
\noalign{\smallskip} \hline \noalign{\smallskip}
1342180683 & $\gamma$~Dra & 3.201 & 0.039 &  &  & 0.585 & 0.076\\
1342182990 & $\gamma$~Dra & 3.271 & 0.030 &  &  & 0.675 & 0.075\\
1342182991 & $\gamma$~Dra & 3.279 & 0.039 &  &  & 0.628 & 0.076\\
1342182993 & $\gamma$~Dra & 3.257 & 0.034 &  &  & 0.660 & 0.074\\
1342182994 & $\gamma$~Dra &  &  & 1.589 & 0.027 & 0.603 & 0.090\\
1342182995 & $\gamma$~Dra &  &  & 1.563 & 0.030 & 0.662 & 0.083\\
1342182996 & $\gamma$~Dra &  &  & 1.586 & 0.029 & 0.674 & 0.065\\
1342186140 & $\gamma$~Dra & 3.301 & 0.034 &  &  & 0.636 & 0.075\\
1342186191 & $\gamma$~Dra & 3.314 & 0.032 &  &  & 0.614 & 0.070\\
1342188069 & $\gamma$~Dra & 3.313 & 0.041 &  &  & 0.593 & 0.078\\
1342189188 & $\gamma$~Dra & 3.286 & 0.036 &  &  & 0.555 & 0.069\\
1342189775 & $\gamma$~Dra & 3.119 & 0.034 &  &  & 0.566 & 0.073\\
1342191124 & $\gamma$~Dra & 3.236 & 0.034 &  &  & 0.630 & 0.084\\
1342191848 & $\gamma$~Dra & 3.261 & 0.036 &  &  & 0.632 & 0.103\\
1342191957 & $\gamma$~Dra & 3.189 & 0.036 &  &  & 0.593 & 0.091\\
1342191960 & $\gamma$~Dra &  &  & 1.568 & 0.027 & 0.603 & 0.073\\
1342192779 & $\gamma$~Dra & 3.217 & 0.039 &  &  & 0.630 & 0.099\\
1342195482 & $\gamma$~Dra & 3.228 & 0.035 &  &  & 0.617 & 0.071\\
1342196729 & $\gamma$~Dra & 3.207 & 0.031 &  &  & 0.664 & 0.080\\
1342198498 & $\gamma$~Dra & 3.221 & 0.035 &  &  & 0.622 & 0.093\\
1342199480 & $\gamma$~Dra & 3.229 & 0.039 &  &  & 0.578 & 0.076\\
1342199511 & $\gamma$~Dra & 3.270 & 0.038 &  &  & 0.574 & 0.080\\
1342199525 & $\gamma$~Dra & 3.271 & 0.038 &  &  & 0.573 & 0.061\\
1342199599 & $\gamma$~Dra & 3.273 & 0.035 &  &  & 0.601 & 0.093\\
1342199638 & $\gamma$~Dra & 3.271 & 0.035 &  &  & 0.614 & 0.055\\
1342199654 & $\gamma$~Dra & 3.289 & 0.041 &  &  & 0.620 & 0.080\\
1342199706 & $\gamma$~Dra & 3.277 & 0.033 &  &  & 0.601 & 0.073\\
1342199716 & $\gamma$~Dra & 3.284 & 0.033 &  &  & 0.615 & 0.072\\
1342202941 & $\gamma$~Dra & 3.269 & 0.036 &  &  & 0.649 & 0.068\\
1342204208 & $\gamma$~Dra & 3.277 & 0.033 &  &  & 0.665 & 0.074\\
1342206000 & $\gamma$~Dra & 3.259 & 0.034 &  &  & 0.626 & 0.062\\
\noalign{\smallskip}\hline
\end{tabular}
\end{table}

\addtocounter{table}{-1}
\begin{table}[!h]
\caption{continued.}
\begin{tabular}{ll@{\hspace{1cm}}cc@{\hspace{1cm}}cc@{\hspace{1cm}}cc}
\hline \noalign{\smallskip}
 & & \multicolumn{6}{c}{Measured flux with correction applied [Jy]} \\
\multicolumn{1}{c}{OBSID} & \multicolumn{1}{c}{Object} & 
$F_{70}$ & $\sigma_{70}$ &
$F_{100}$ & $\sigma_{100}$ &
$F_{160}$ & $\sigma_{160}$ \\
\noalign{\smallskip} \hline \noalign{\smallskip}
1342208970 & $\gamma$~Dra & 3.263 & 0.037 &  &  & 0.608 & 0.068\\
1342210581 & $\gamma$~Dra & 3.242 & 0.037 &  &  & 0.616 & 0.066\\
1342212493 & $\gamma$~Dra & 3.257 & 0.033 &  &  & 0.527 & 0.086\\
1342212496 & $\gamma$~Dra &  &  & 1.598 & 0.027 & 0.566 & 0.080\\
1342213587 & $\gamma$~Dra & 3.215 & 0.040 &  &  & 0.650 & 0.066\\
1342215373 & $\gamma$~Dra & 3.227 & 0.038 &  &  & 0.621 & 0.069\\
1342216068 & $\gamma$~Dra &  &  & 1.578 & 0.027 & 0.621 & 0.055\\
1342217403 & $\gamma$~Dra & 3.221 & 0.036 &  &  & 0.621 & 0.085\\
1342220822 & $\gamma$~Dra & 3.291 & 0.032 &  &  & 0.628 & 0.087\\
1342221810 & $\gamma$~Dra & 3.259 & 0.033 &  &  & 0.684 & 0.067\\
1342222755 & $\gamma$~Dra & 3.301 & 0.036 &  &  & 0.603 & 0.079\\
1342224228 & $\gamma$~Dra & 3.288 & 0.035 &  &  & 0.619 & 0.071\\
1342226711 & $\gamma$~Dra & 3.225 & 0.034 &  &  & 0.669 & 0.074\\
1342228387 & $\gamma$~Dra & 3.255 & 0.035 &  &  & 0.650 & 0.066\\
1342228390 & $\gamma$~Dra &  &  & 1.569 & 0.027 & 0.638 & 0.081\\
1342231096 & $\gamma$~Dra & 3.232 & 0.037 &  &  & 0.609 & 0.076\\
1342231898 & $\gamma$~Dra & 3.227 & 0.034 &  &  & 0.613 & 0.064\\
1342234213 & $\gamma$~Dra & 3.247 & 0.033 &  &  & 0.636 & 0.068\\
1342237974 & $\gamma$~Dra & 3.238 & 0.038 &  &  & 0.596 & 0.067\\
1342238771 & $\gamma$~Dra & 3.262 & 0.030 &  &  & 0.652 & 0.083\\
1342240698 & $\gamma$~Dra & 3.249 & 0.034 &  &  & 0.638 & 0.073\\
1342242556 & $\gamma$~Dra & 3.264 & 0.031 &  &  & 0.590 & 0.077\\
1342244899 & $\gamma$~Dra & 3.258 & 0.038 &  &  & 0.638 & 0.070\\
1342246180 & $\gamma$~Dra & 3.269 & 0.033 &  &  & 0.571 & 0.073\\
1342247334 & $\gamma$~Dra & 3.236 & 0.032 &  &  & 0.608 & 0.067\\
1342247337 & $\gamma$~Dra &  &  & 1.583 & 0.025 & 0.602 & 0.071\\
1342248037 & $\gamma$~Dra & 3.292 & 0.034 &  &  & 0.629 & 0.075\\
1342249292 & $\gamma$~Dra & 3.289 & 0.033 &  &  & 0.631 & 0.068\\
1342250855 & $\gamma$~Dra & 3.272 & 0.035 &  &  & 0.632 & 0.088\\
1342252804 & $\gamma$~Dra & 3.239 & 0.034 &  &  & 0.624 & 0.059\\
1342254722 & $\gamma$~Dra & 3.244 & 0.035 &  &  & 0.642 & 0.060\\
1342256958 & $\gamma$~Dra & 3.316 & 0.035 &  &  & 0.601 & 0.064\\
1342258830 & $\gamma$~Dra & 3.266 & 0.036 &  &  & 0.638 & 0.069\\
1342258833 & $\gamma$~Dra &  &  & 1.595 & 0.027 & 0.591 & 0.081\\
1342262224 & $\gamma$~Dra & 3.255 & 0.031 &  &  & 0.608 & 0.074\\
1342267290 & $\gamma$~Dra & 3.228 & 0.036 &  &  & 0.679 & 0.096\\
1342268965 & $\gamma$~Dra & 3.226 & 0.031 &  &  & 0.575 & 0.070\\
1342269811 & $\gamma$~Dra & 3.251 & 0.033 &  &  & 0.645 & 0.088\\
1342270999 & $\gamma$~Dra & 3.212 & 0.032 &  &  & 0.599 & 0.087\\
\noalign{\smallskip}\hline
\end{tabular}
\end{table}

\clearpage
\section{HIPE tasks}
\label{s:tasks}

\subsection{Flux correction}
\label{s:fluxcorr}

{\scriptsize
\begin{verbatim}
"""
fluxCalCorrPs2Scan()
====================
This task scales the flux determined from aperture photometry of chopped PS
observations to the flux calibration scheme of scan maps. The conversion is
valid for integrated flux of up to approximately 10 Jy only.

Input parameters:
frames: the frames class
ft: time of the observation in FineTime
    (e.g. using getObstimeFromFrames) 

Output parameters:
smframes: frames scaled to scan map flux scale
"""
def fluxCalCorrPs2Scan(frames,ft):
    pstosmcorrs = TableDataset(description = "Chopped phot flux correction")
    pstosmcorrs["Filter"]=Column(String1d(["blue","green","red"]))
    pstosmcorrs["C1"]=Column(Double1d([0.924,0.937,0.949]))
    pstosmcorrs["C2"]=Column(Double1d([6.551e-4,4.276e-4,1.948e-4]))

    smframes = frames.copy()
    if (smframes.meta["type"].value == "HPPAVGR"):
        filter = "red"
    elif (smframes.meta["blue"].value == "blue1"):
        filter = "blue"
    else:
        filter = "green"
    row = pstosmcorrs["Filter"].data.where(pstosmcorrs["Filter"].data == 
filter).toInt1d()
    c1 = pstosmcorrs["C1"].data[row]
    c2 = pstosmcorrs["C2"].data[row]
    corr = c1+c2/(ft.microsecondsSince1958()*1.0e-15-1.62)
    for i in range(smframes.getNumberOfFrames()):
        signal = smframes.refs[i].product["Signal"].data/corr
        smframes.refs[i].product["Signal"].data = signal
    pass
    return smframes
\end{verbatim}
}

\subsection{Time of observation in \texttt{FineTime}}
\label{s:ft}

{\scriptsize
\begin{verbatim}
"""
getObstimeFromFrames()
======================
This task extracts the actual time at the middle of the observation. Start and
end times are extracted from level 0 detector data of the Observation Context.
The central time is returned as a FineTime.

Input parameter:
obs: Observation Context of the observation
"""
def getObstimeFromFrames(obs):
    frames = obs.level0.refs["HPPAVGB"].product
    framesStart = frames.getStartDate().microsecondsSince1958()
    framesEnd = frames.getEndDate().microsecondsSince1958()
    framesTime = (framesStart+framesEnd)/2
    ft=FineTime(framesTime)
    return ft    
\end{verbatim}
}

\end{document}